\documentclass[a4paper,11pt]{article}
\pdfoutput=1 % if your are submitting a pdflatex (i.e. if you have
\usepackage{jheppub} % for details on the use of the package, please
\usepackage{amsmath,amsthm,physics,mathtools}
\usepackage[T1]{fontenc} % if needed

\newcommand{\OH}{{\cal O}_H}

\newcommand{\OO}{{\cal O}}

\newcommand{\ra}{{\rightarrow}}

\title{Thermal Stress Tensor Correlators, OPE and Holography}

\author[a]{Robin Karlsson,}
\author[a]{Andrei Parnachev,}
\author[b]{Valentina Prilepina,}
\author[a]{Samuel Valach}

\affiliation[a]{School of Mathematics and Hamilton Mathematics Institute,\setlength{\parskip}{0pt}

Trinity College Dublin, Dublin 2, Ireland}
\affiliation[b]{Stony Brook University, Stony Brook NY, USA}

\emailAdd{karlsson@maths.tcd.ie}
\emailAdd{parnachev@maths.tcd.ie}
\emailAdd{valentina.prilepina.1@ulaval.ca}
\emailAdd{valachs@tcd.ie}

\abstract{In strongly coupled conformal field theories with a large central charge important light degrees of freedom are the stress tensor and its composites, multi-stress tensors. We consider the OPE expansion of two-point functions of the stress tensor in thermal and heavy states and focus on the contributions from the stress tensor and double-stress tensors in four spacetime dimensions. We compare the results to the holographic finite temperature two-point functions and read off  conformal data beyond the leading order in the large central charge expansion.	In particular, we compute corrections to the OPE coefficients which determine the near-lightcone behavior of the  correlators. We also compute  the anomalous dimensions of the double-stress tensor operators.}

\begin{document} 
\maketitle
\flushbottom

\section{Introduction and summary}

Hydrodynamics describes low-energy excitations in matter at finite temperature and density \cite{Landau1987}.
A lot of interest was attracted to the hydrodynamics of conformal field theories at strong coupling and large central charge $C_T$,
which  admit a dual
gravitational (holographic) description \cite{Maldacena:1997re,Gubser:1998bc,Witten:1998qj}.
Transport coefficients  can be extracted from the two-point functions of the stress tensor (TT-correlators) at finite temperature
and holography maps these correlators to two-point functions of metric perturbations in a black hole background
 \cite{Policastro:2001yc,Policastro:2002se,Kovtun:2003wp,Kovtun:2004de}.

The holographic value of the shear viscosity is much closer to the experimentally observed values for quark-gluon plasma
than perturbative calculations  (see e.g. \cite{Teaney:2009qa,Heinz:2013th}  for  reviews).
The ratio of the shear viscosity to the entropy density was shown to be universal,
$\eta/s =\hbar/4\pi k_B$,  in all theories
with  Einstein gravity duals \cite{Kovtun:2003wp,Kovtun:2004de,Buchel:2003tz,Iqbal:2008by}.
However, the addition of higher derivative terms to the gravity Lagrangian 
changes this  value  \cite{Kats:2007mq,Brigante:2007nu,Brigante:2008gz}. What does this imply for the hydrodynamics of strongly interacting
field theories?

In a way, gravity provides a minimal model for strongly interacting matter, where the only degrees of freedom 
are  the stress tensor and its composites, multi-stress tensors  -- they are encoded by the fluctuations of the metric in the dual theory.
From a CFT point of view, such a minimal model is defined by the OPE coefficients and the spectrum of anomalous dimensions of
multi-stress tensor operators. 
Consider the OPE coefficients which determine the three-point functions of the stress tensor, 
which are specified by the three parameters  in $d>3$ dimensions.
They change as the bulk couplings in front of the gravitational higher derivative terms are varied.\footnote{
Note that we expect consistent holographic models with generic graviton three-point couplings to also contain
higher spin fields \cite{Camanho:2014apa}. }
Presumably these OPE coefficients do not completely determine the theory, but is it possible that some sector of the theory 
is universal?

We can make progress in answering this question by decomposing the TT correlator using the OPE expansion. In a minimal theory
the operators that appear are multi-stress tensor operators\footnote{
In this paper we consider Einstein gravity as a holographic model -- it is believed to be a consistent 
truncation. In other words, in the dual CFT language, couplings to other operators and corrections to the OPE
coefficients are suppressed by the (large) gap in the spectrum of the conformal dimensions of higher spin operators --
see e.g. \cite{Meltzer:2017rtf,Afkhami-Jeddi:2018own} for  recent discussions.}
 and one can in principle deduce the conformal data working order-by-order
in the temperature $T=\beta^{-1}$  [in $d$ spacetime dimensions $k$-stress tensors naturally contribute terms $\OO(\beta^{-dk})$]. 
A similar question was recently asked in a simpler setting where a finite temperature state (dual to a black hole)  was  probed by scalars \cite{Fitzpatrick:2019zqz}.
A scalar  two-point function has a piece which can be computed near the boundary of asymptotically AdS spacetime -- this is precisely
the term which encodes the contribution of multi-stress tensors.
Another piece, left undetermined in the near-boundary expansion, contains the contributions of multi-trace operators which 
involve the external scalar operator.
To compute it, one needs to solve the equation of motion in the whole spacetime -- a nontrivial task in practice.
\footnote{In \cite{Karlsson:2019dbd} an alternative way of computing the stress tensor sector of the scalar correlator using conformal bootstrap
	and an ansatz, motivated by \cite{Kulaxizi:2019tkd}, was proposed. 
	The procedure of \cite{Karlsson:2019dbd}  allows one to compute the OPE coefficients with the leading twist multi-stress tensors.
	The result has many similarities to the Virasoro HHLL vacuum block (see e.g. \cite{Fitzpatrick:2014vua,Fitzpatrick:2015zha,Kulaxizi:2018dxo,Karlsson:2021mgg})  but at the moment
	the full resummed correlator in $d>2$ is only known in the $\Delta_L \ra \infty$ limit \cite{Parnachev:2020fna}.
	(see \cite{,Karlsson:2019qfi,Li:2019tpf,Huang:2019fog,Fitzpatrick:2019efk,Li:2019zba,Karlsson:2019txu,Huang:2020ycs,Karlsson:2020ghx,Li:2020dqm,Fitzpatrick:2020yjb,Parnachev:2020zbr,Karlsson:2021duj,Rodriguez-Gomez:2021pfh,Huang:2021hye,Rodriguez-Gomez:2021mkk,Krishna:2021fus,Huang:2022vcs,Dodelson:2022eiz} for related work).}

What happens when a thermal state (or, in the dual language, a black hole) is probed by the stress tensor operators?
In this paper we attempt to decompose this correlator by
generalizing the analysis of   \cite{Fitzpatrick:2019zqz} to the case of external operator being the stress tensor.
Here we consider the contributions of the identity operator, the stress tensor and the double stress tensors
to the correlator.
One immediate technical complication that we face is that the external operator with which we probe the system, namely the
stress tensor, has integer conformal dimension.
In   \cite{Fitzpatrick:2019zqz}  it was observed that some OPE coefficients have poles for integer values of the conformal dimension of the external scalar operator.
This feature is related to mixing of double stress and double trace operators.
The OPE coefficients for both series have poles which cancel,
leaving behind logarithmic terms.
One can also observe that the coefficients of these terms cannot be fixed by the near-boundary analysis \cite{Fitzpatrick:2019zqz,Li:2019tpf}.

In the case of the stress tensor the double-trace operators made out of the external operator $T_{\mu\nu}$ {\it are also}  double stress tensor operators.
One may wonder if their OPE coefficients can be determined from the near boundary analysis.
The answer turns out to be no.
Another important difference from  \cite{Fitzpatrick:2019zqz}  is related to the leading behavior of the OPE coefficients of two
stress tensors and a double stress tensor.
This OPE coefficient scales like one (for unit-normalized operators), as opposed to $\OO(1/C_T)$ in the scalar case, and gives rise to the disconnected
part of  the correlator.
This implies that the connected part of the TT correlator contains information about conformal data which is subleading in the
$1/C_T$ expansion.
This leads to some complications\footnote{
In particular, it has been observed  in the $d=2$ case that thermalization of multi stress-tensor operators happens only
to leading order in the $1/C_T$ expansion (see \cite{Basu:2017kzo,He:2017txy,Datta:2019jeo} for recent discussions).}
 but in the end, we succeed at extracting the leading $1/C_T$ contributions to the anomalous dimensions of the double trace operators
and to the leading lightcone behavior of the TT correlators.
Other conformal data at this order remains undetermined -- it should be thought of as an analog of the double trace operator data in the 
external scalar case.

Let us mention another technical difficulty that we need to confront in the case of external stress tensors.
In \cite{Fitzpatrick:2019zqz}  the symmetry implies that the two-point function depends on the time $t$, the spatial radial coordinate $\rho$ and 
the AdS radial coordinate $r$.
This is no longer true in the stress tensor case, due to the presence of distinct polarizations.
We handle this by computing stress tensor correlators integrated over two parallel $xy$-planes separated in the transverse spatial 
direction, which we denote by $z$. 
There are three independent choices of polarization, distinguished by the  transformation properties with respect
to rotations of the plane of integration.
A suitable modification of the ansatz used in  \cite{Fitzpatrick:2019zqz} allows us to solve the stress tensor problem.
However, integrating over the $xy$-plane leads to some divergent contributions and to additional logarithmic terms.
Fortunately, this does not affect our ability to extract the anomalous dimensions.

The rest of this paper is organized as follows. In Section \ref{Section:Bulk}, we consider metric perturbations on top of a planar AdS-Schwarzschild black hole and compute the stress tensor two-point function in a near-boundary expansion (OPE limit in the dual CFT). In Section \ref{CFTside4d}, we perform the OPE expansion of the stress tensor thermal two-point functions in $d=4$ and by comparison to the bulk calculations in the previous section, we read off the anomalous dimensions of double-stress tensor operators with spin $J=0,2,4$
and determine the near-lightcone behavior of the correlators. We conclude with a discussion in Section \ref{sec:disc}. 
In Appendix \ref{AppendixA}, we treat the simpler example of scalar perturbations in the bulk as a toy model for the metric perturbations, 
focusing  on the subtleties that arise for external operators with integer dimensions.
In addition, we consider scalar correlators integrated over the $xy$-plane and 
show  how the correct OPE data is recovered in this case. 
Appendix \ref{AppendixB} lists some of the results that are too lengthy to present in Section \ref{Section:Bulk}.
 In Appendix \ref{app:SpinningBlocks}, we present conventions and details on the spinning conformal correlators
 relevant for the decomposition of  thermal stress tensor two-point functions.

\section{Holographic calculation of thermal  TT correlators}\label{Section:Bulk}

Recently some OPE coefficients of scalars and multi-stress tensors  were calculated in 
the context of holographic models \cite{Fitzpatrick:2019zqz,Li:2019tpf}. 
This was accomplished by making a  comparison 
between the CFT conformal block decomposition of  HHLL correlators on the CFT side 
and a near-boundary expansion of
the bulk-to-boundary propagator in the AdS-Schwarzschild background on the bulk side.

Our goal in this work is to use an analogous approach to extract the CFT data\footnote{By the CFT data we mean products of the OPE coefficients and thermal one-point functions and anomalous dimensions of the double-trace stress tensors. This will be explained in greater detail in the next section.} for the stress tensor two-point function in a thermal state dual to the AdS-Schwarzschild black hole, in this section we will focus on the bulk part of this calculation. In practice we will consider the integrated version of the correlator
\begin{equation}\label{eq:defHTTH}
	G_{\mu\nu,\rho\sigma}(t,z):=\int_{\mathbb{R}^2} dx dy\langle T_{\mu\nu}(x^\alpha)T_{\rho\sigma}(0)\rangle_{\beta}.
\end{equation}

To compute the TT correlator, it is necessary to consider the linearized Einstein equations in the black hole background.
For technical reasons, we will take the large volume limit, where all conformal descendants decouple and
an expansion in terms of conformal blocks becomes  the OPE expansion.
On the bulk side, this corresponds to considering the planar asymptotically AdS  black hole. 
The corresponding system of PDEs is technically difficult to solve because different polarizations mix with each other. 
To make the problem tractable, we integrate the correlator over two spatial directions in \eqref{eq:defHTTH}.
The resulting fluctuation equations simplify to  three independent PDEs for the three different polarizations.
We   show explicitly that  an  ansatz of  \cite{Fitzpatrick:2019zqz,Li:2019tpf}, suitably modified to fit our needs, 
successfully solves these equations.

As a warm-up exercise, we consider the scalar case, discussed in \cite{Fitzpatrick:2019zqz,Li:2019tpf}, but now integrate 
over the $xy$-plane.
The details of this calculation are described in Appendix \ref{AppendixA}, but the summary is as follows.
For non-integer values $\Delta_L$  of the conformal dimensions of the scalar operator
all coefficients in the ansatz are fixed, order-by-order,
by imposing the scalar field equations of motion in the bulk.
Matching to the conformal block expansion then yields the OPE coefficients of scalars and multi-stress tensors,
which reproduce the results of \cite{Fitzpatrick:2019zqz}.
Note that the integrals are only convergent for large $\Delta_L$, but their analytic continuation to small  $\Delta_L$ yields the correct results.

For integer $\Delta_L$ there is mixing between multi-stress and multi-trace operators, which results in logarithmic terms \cite{Fitzpatrick:2019zqz}.
This mixing is reflected in the appearance of the $\log r$ terms in the bulk ansatz \cite{Li:2019tpf}; a closely related fact is 
that not all coefficients in the ansatz are now determined by the bulk equations of motion.
For example, for $\Delta_L=4$  there is one undetermined parameter at $\OO(\mu^2)$; it corresponds to
an undetermined factor in a double-trace OPE coefficient.

As explained in Appendix A, the addition of spatial integration leads to an additional undetermined coefficient in the ansatz. 
This  coefficient is, roughly speaking, related to the volume of the $xy$-plane we are integrating over.
In practice, we use dimensional regularization, so instead of the volume, a $1/\epsilon$ pole appears in the expression for
this undetermined coefficient.
The other undetermined coefficient is related to the logarithmic term, just as in the non-integrated case.
In summary, we conclude that in the  scalar case, the spatial integration does not affect our ability to read off the OPE data.

In this section we perform the bulk calculations for the case where the external operator is the stress tensor. 
In other words, we compute the OPE expansion for the thermal TT correlator in holographic CFTs.
This section is organized as follows. First we consider metric perturbations around a planar AdS-Schwarzschild black hole. Then we integrate out two out of five space-time directions and, following \cite{Kovtun:2005ev,Kovtun:2006pf}, we utilize the resulting $O(2)$ symmetry together with the bulk gauge freedom to reformulate the problem in terms of the 
three gauge invariant combinations of the gravitational fluctuations in the AdS-Schwarzschild background.
The resulting PDEs can then be solved one by one using the ansatz \cite{Fitzpatrick:2019zqz,Li:2019tpf}, naturally adapted to the integrated case. Finally, using the holographic dictionary, we derive the stress tensor two-point function in a thermal state for various polarizations. In Section \ref{CFTside4d} we compare these results with the CFT conformal block decomposition and extract conformal data.  

\subsection{Linearized Einstein equations}

We consider the Einstein-Hilbert action with a cosmological constant\footnote{We will be using the Euclidean signature throughout.}
\begin{equation}\label{EHAction}
	S=\frac{1}{16\pi G_5}\int\dd^5x\sqrt{g}(\mathcal{R}-2\Lambda),
\end{equation}
where $G_5$ is the five-dimensional gravitational constant, $\mathcal{R}$ is the Ricci scalar and $\Lambda$ is the cosmological constant. Decomposing the metric as the background part plus a small perturbation $h_{\mu\nu}$, one obtains the linearized Einstein equations
in the form
\begin{equation}\label{LEE}
	R^{(1)}_{\mu\nu}+dh_{\mu\nu}=0,
\end{equation}
where $R_{\mu\nu}^{(1)}$ is the linearized Ricci tensor and $d$ is the dimension of the conformal boundary, i.e. $d=4$ in our case. 

We will be interested in the planar AdS-Schwarzschild black hole as the background spacetime,
\begin{equation}\label{bhmetrics}
	\dd s^2 = r^2f(r)\dd t^2+r^2\dd \vec{x}^2+\frac{1}{r^2f(r)}\dd r^2,
\end{equation}
where $\vec{x}=(x,y,z)$ and $f(r)=1-\frac{\mu}{r^4}$.
Here and in the rest of the paper we set the radius of the AdS space to unity.

By solving the linearized Einstein equations (\ref{LEE}) with the appropriate boundary conditions, we obtain the metric perturbation $h_{\mu\nu}$ and, in principle, the holographic dictionary then precisely determines the correlators in the four-dimensional CFT on the boundary. However, due to the complicated form of these equations, this is difficult to do in practice.

To make this problem tractable, we integrate the bulk-to-boundary propagator over the $xy$-plane. This will simplify the equations of motion to three independent PDEs, which we will be able to solve using the ansatz \cite{Fitzpatrick:2019zqz,Li:2019tpf}.
As a result, the corresponding CFT correlators, which  we obtain via holographic dictionary, will be integrated over 
the $xy$ directions. This will be  studied in Section \ref{CFTside4d} from the  CFT point of view.

\subsection{Polarizations and gauge invariants}

Our aim is to solve the linearized Einstein equations \eqref{LEE} in the background \eqref{bhmetrics}, with the solution integrated over two spatial directions, which we choose to be $x$ and $y$.

Upon  integration, the (linearized) gravitational action will exhibit an $O(2)$ rotational symmetry. This property allows us to divide the components $h_{\mu\nu}$ into three representations (referred to as channels in this context) which can be studied separately:
\begin{align}
	\text{Sound channel}&:\qquad h_{tt},\,h_{tz},\,h_{zz},\,h_{rr},\,h_{tr},\,h_{zr},\,h_{xx}+h_{yy}\\
	\text{Shear channel}&:\qquad h_{tx},\,h_{ty},\,h_{zx},\,h_{zy},\,h_{rx},\,h_{ry}\\
	\text{Scalar channel}&:\qquad h_{\alpha\beta}-\delta_{\alpha\beta}(h_{xx}+h_{yy})/2.
\end{align}
The sound channel has spin 0, shear channel has spin 1 and the scalar channel (whose equations of motion will be identical 
to that of the scalar) has spin 2 under $O(2)$.

In every channel, we can define a quantity $Z_i$ \cite{Kovtun:2005ev,Kovtun:2006pf}, that is invariant under the gauge transformations $h_{\mu\nu}\rightarrow h_{\mu\nu}-\nabla_\mu\xi_\nu-\nabla_\nu\xi_\mu$ of the gravitational bulk theory. In the position space these are
\begin{align}
	Z_1&=\partial_zH_{tx}-\partial_tH_{xz}\label{vectorinvdef}\\
	Z_2&=2f\partial_z^2H_{tt}-4\partial_t\partial_zH_{tz}+2\partial_t^2H_{zz}-\left((f+\frac{r}{2}f')\partial_z^2+\partial_t^2\right)(H_{xx}+H_{yy})\label{soundinvdef}\\
	Z_3&=H_{xy}\label{tensorinvdef},
\end{align}
where $H_{tt}=h_{tt}/fr^2$, $H_{ti}=h_{ti}/r^2$ and $H_{ij}=h_{ij}/r^2$ for $i,j\in\{x,y,z\}$, $f=f(r)$ is the function appearing in the black hole metric and the prime denotes the derivative with respect to $r$. 
As is conventional, we refer to  $Z_1$, $Z_2$  and  $Z_3$ as  the shear channel invariant, the sound channel invariant and the scalar channel
invariant, respectively.

We can now choose a particular channel, take the linearized Einstein equations \eqref{LEE} and assume the metric perturbation 
to be of the form $h_{\mu\nu}=h_{\mu\nu}(t,z,r)$. Combining the resulting equations, we get PDEs for the invariants. 
It will be useful to define the following quantities:
\begin{align}
	c_1&\coloneqq (3\mu^2-8\mu r^4+5r^8)/r^5\\
	c_2&\coloneqq 2\mu(r^4-\mu)/r^5\\
	c_3&\coloneqq (\mu-r^4)^2/r^4\\
	c_4&\coloneqq 16\mu^2(r^4-\mu)/(3r^{10})\\
	c_5&\coloneqq 1+\mu(\mu-4r^4)/(3r^8)\\
	c_6&\coloneqq 2-4\mu/(3r^4)\\
	c_7&\coloneqq (\mu^2-6\mu r^4+5r^8)/r^5\\
	c_8&\coloneqq (r^4-\mu)(9\mu^2-16\mu r^4+15r^8)/(3r^9)\\
	c_9&\coloneqq -(\mu-3r^4)(\mu-r^4)^2/(3r^8).
\end{align}
The equations of motion for the invariants are then given by\footnote{These are the equations one finds by Wick rotating and Fourier transforming the corresponding ODEs presented in \cite{Kovtun:2005ev}.}
\begin{align}
	0&=(\partial_t^2+f\partial_z^2)^2Z_1+\left(c_1(\partial_t^2+f\partial_z^2)+c_2(\partial_t^2-f\partial_z^2)\right)Z_1'+c_3(\partial_t^2+f\partial_z^2)Z_1''\label{es1}\\
	0&=(c_4\partial_z^2+c_5\partial_z^4+c_6\partial_t^2\partial_z^2+\partial_t^4)Z_2+(c_7\partial_t^2+c_8\partial_z^2)Z_2'+(c_3\partial_t^2+c_9\partial_z^2)Z_2''\label{es2}\\
	0&=(\partial_t^2+f\partial_z^2)Z_3+c_7Z_3'+c_3Z_3''.\label{es3}
\end{align}

\subsection{Ansatz and the vacuum propagators}\label{susucica}

In order to solve (\ref{es1})-(\ref{es3}) we need to find the bulk-to-boundary propagators $\mathcal{Z}_i$, which are related to the invariants by
\begin{equation}\label{prevodis}
	Z_i(t,z,r)=\int\dd t'\dd z'\mathcal{Z}_i(t-t',z-z',r)\hat{Z}_i(t',z'),
\end{equation}
where $\hat{Z}_i$ is related to the boundary value (up to derivatives) of $Z_i$ as will be explained below. To solve the equations of motion we use the ansatz \cite{Fitzpatrick:2019zqz, Li:2019tpf} introduced for the case of a scalar field in a black hole background, suitably modified for our integrated case. Let us briefly review its derivation and the logic behind its construction.

Although in $d=4$ the bulk equations cannot be solved analytically, one can try to find an expansion of the solution corresponding to the OPE limit on the boundary and extract the CFT data. As was successfully demonstrated in \cite{Fitzpatrick:2019zqz, Li:2019tpf}, such a bulk regime is given by\footnote{Note that in the original non-integrated case one has $|\vec{x}|$ instead of $z$.}
\begin{equation}\label{limitoss}
	r\rightarrow\infty\qquad\text{with}\qquad rt,\, rz\,\,\,\text{fixed}
\end{equation}
The intuition behind this limit is the expectation that the bulk solution becomes sensitive only to the near-boundary region as the CFT operators approach each other.

To realize \eqref{limitoss} in practice, it is useful to introduce new coordinates defined by
\begin{align}
	\rho&\coloneqq rz\\
	w^2&\coloneqq 1+r^2t^2+r^2z^2\,.
\end{align}
In these coordinates the limit is $r\rightarrow\infty$ with $w$ and $\rho$ held fixed. 

According to \cite{Fitzpatrick:2019zqz}, one expects the solution to be of the form of the product of 
 the AdS propagator and  an  expansion in $1/r$, where at each order we have a polynomial $\sum_i\alpha_i(w)\rho^i$. Substituting this into the equations of motion, we can find analytical solutions for all $\alpha_i(w)$.
 Imposing regularity in the bulk and demanding the proper boundary behaviour\footnote{By the proper boundary behaviour we mean that the boundary limit of the bulk solution should reproduce the form of the boundary correlators expected from the boundary CFT.}, we determine the integration constants and find the coefficients $\alpha_i(w)$ as polynomials in $w$.

If there are logarithmic terms\footnote{Logarithmic terms appear, for example, in the case of a scalar field with integer conformal dimension $\Delta_L$ or in the presence of anomalous dimensions as in the case of the stress tensor thermal two-point function. They can  also be produced upon integration. We comment more on the origin of these terms in appendix \ref{AppendixA}.}
$\mathcal{Z}_i$ takes the form \cite{Li:2019tpf}
\begin{equation}\label{thetheansatz}
	\mathcal{Z}_i=\mathcal{Z}_i^{AdS}\left(1+\frac{1}{r^4}\left(G^{4,1}_i+G^{4,2}_i\log r\right)+\frac{1}{r^8}\left(G^{8,1}_i+G^{8,2}_i\log r\right)+\ldots\right),    
\end{equation}
where $\mathcal{Z}_i^{AdS}$ is the vacuum bulk-to-boundary propagator for the invariant $Z_i$ and $G^{4,j}_i$, $G^{8,j}_i$, $\ldots$, $j\in\{1,2\}$, $i\in\{1,2,3\}$ are given by\footnote{We use the bounds of the sums as they were derived for the case of a scalar field \cite{Fitzpatrick:2019zqz,Li:2019tpf}. As we will see, this will be valid also for the stress tensor case in the scalar and shear channels. In the sound channel we will need a slight modification of the ansatz.} (we suppress the channel index for simplicity)
\begin{align}
	&G^{4,j}=\sum_{m=0}^2\sum_{n=-2}^{4-m}(a^{4,j}_{n,m}+b^{4,j}_{n,m}\log w)w^n\rho^m\\
	&G^{8,j}=\sum_{m=0}^6\sum_{n=-6}^{8-m}(a^{8,j}_{n,m}+b^{8,j}_{n,m}\log w)w^n\rho^m\\
	&\phantom{wwwwww}\vdots\nonumber
\end{align}
Here $G^{4,j}$ corresponds to the stress tensor contribution ($\propto\mu^1$) and $G^{8,j}$ corresponds to the double-stress tensor contributions. We expect to find $b^{4,1}=0$ and $G^{4,2}=0$ in all three channels. 

The vacuum propagator $\mathcal{Z}_i^{AdS}$ for the $i$-th channel can be determined using the AdS bulk-to-boundary propagators for the various components of the metric perturbation.
Let us describe this calculation in more detail. The AdS propagator for $H_{\mu\nu}$ was computed in \cite{Liu:1998bu} and in the five dimensional bulk case can be expressed as 
\begin{equation}
	\mathfrak{G}_{\mu\nu,\rho\sigma}=\frac{10r^4}{\pi^2(1+r^2(t^2+\vec{x}^2))^4}J_{\mu\alpha}J_{\nu\beta}P_{\alpha\beta,\rho\sigma},
\end{equation}
where $J_{\mu\nu}$ and $P_{\mu\nu,\rho\sigma}$ are given by
\begin{align}
	J_{\mu\nu}&=\delta_{\mu\nu}-\frac{2x_\mu x_\nu}{\frac{1}{r^2}+t^2+\vec{x}^2}\\
	P_{\mu\nu,\rho\sigma}&=\frac12(\delta_{\mu\rho}\delta_{\nu\sigma}+\delta_{\nu\rho}\delta_{\mu\sigma})-\frac14\delta_{\mu\nu}\delta_{\rho\sigma}.
\end{align}
Integrating out the $x$ and $y$ directions, we get
\begin{equation}\label{vacHsgreen}
	\mathcal{G}_{\mu\nu,\rho\sigma}(t,z,r)\coloneqq\int_{\mathbb{R}^2}\dd x\dd y\,\mathfrak{G}_{\mu\nu,\rho\sigma}(t,x,y,z,r),
\end{equation}
from which we find the (integrated) AdS solution for $H_{\mu\nu}$ 
\begin{equation}\label{vacHs}
	H_{\mu\nu}(t,z,r)=\int\dd t'\dd z'\mathcal{G}_{\mu\nu\rho\sigma}(t-t',z-z',r)\hat{H}_{\rho\sigma}(t',z'),
\end{equation}
where $\hat{H}_{\mu\nu}$ are the sources, i.e. the values of the bulk solution on the conformal boundary.

Substituting \eqref{vacHs} into the definitions of the invariants (\ref{vectorinvdef})-(\ref{tensorinvdef}), one can
accordingly read off the AdS bulk-to-boundary propagators $\mathcal{Z}^{AdS}_i$.

Here we list the resulting expressions for some particular choices of the sources:
{\setlength{\tabcolsep}{19pt}
	\renewcommand{\arraystretch}{2.1}%2.2
	\begin{center}
		\begin{tabular}{c|c c}
			Sources     &  $(t,z,r)$-result & $(w,\rho,r)$-result\\
			\hline\hline
			$\hat{H}_{xy}$    &   $\mathcal{Z}_3^{AdS}=\frac{2r^2}{\pi\left(r^2\left(t^2+z^2\right)+1\right)^3}$    &$=\frac{2r^2}{\pi w^6}$\\[1ex]
			\hline
			$\hat{H}_{tx}$    &   $\mathcal{Z}_1^{AdS}=-\frac{12 r^4 z}{\pi\left(r^2\left(t^2+z^2\right)+1\right)^4}$     &$=-\frac{12r^3\rho}{\pi w^8}$\\
			$\hat{H}_{xz}$    &   $\mathcal{Z}_1^{AdS}=\frac{12 r^4 t}{\pi  \left(r^2 \left(t^2+z^2\right)+1\right)^4}$      &$=\frac{12r^3\sqrt{w^2-1-\rho^2}}{\pi w^8}$\\[1ex]
			\hline
			$\hat{H}_{tz}$    &   $\mathcal{Z}_2^{AdS}=-\frac{384 r^6 t z}{\pi  \left(r^2 \left(t^2+z^2\right)+1\right)^5}$  &$=-\frac{384r^4\rho\sqrt{w^2-1-\rho^2}}{\pi w^{10}}$\\
			$\hat{H}_{tt}$    &   $\mathcal{Z}_2^{AdS}=-\frac{24 \left(r^6 \left(t^2-7 z^2\right)+r^4\right)}{\pi  \left(r^2 \left(t^2+z^2\right)+1\right)^5}$  &$=-\frac{24r^4(w^2-8\rho^2)}{\pi w^{10}}$\\
			$\hat{H}_{xx}$   &   $\mathcal{Z}_2^{AdS}=\frac{24 r^4-72 r^6 \left(t^2+z^2\right)}{\pi  \left(r^2 \left(t^2+z^2\right)+1\right)^5}$  &$=\frac{24r^4(4-3w^2)}{\pi w^{10}}$\\  
			$\hat{H}_{zz}$    &   $\mathcal{Z}_2^{AdS}=\frac{24 r^4 \left(r^2 \left(7 t^2-z^2\right)-1\right)}{\pi  \left(r^2 \left(t^2+z^2\right)+1\right)^5}$  &$=\frac{24r^4(7w^2-8(1+\rho^2))}{\pi w^{10}}$
		\end{tabular}
\end{center}}

At this point, we have all the pieces needed for the ansatz (\ref{thetheansatz}). 
Inserting it into equations (\ref{es1})-(\ref{es3}), we can determine the coefficients $a_{n,m}^{k,j}$ and $b_{n,m}^{k,j}$. 
We next proceed to discuss the results channel-by-channel.

\subsection{Scalar channel}\label{sekskalal}

We begin by considering the scalar channel where the equation of motion (\ref{es3}) has the simplest form. In this paper we confine our attention to the contributions due to the identity operator ($\mu^0$), the stress tensor ($\mu^1$) and double-stress tensors ($\mu^2$). We are therefore interested in finding $G^{4,1}_i$, $G^{4,2}_i$, $G^{8,1}_i$ and $G^{8,2}_i$ in the ansatz (\ref{thetheansatz}). 

In the scalar channel, we may either turn on the source $\hat{H}_{xy}\neq0$ or $\hat{H}_{xx}=-\hat{H}_{yy}\neq0$. Since these differ only by an $O(2)$ rotation, the corresponding bulk solutions, as well as the form of the action will be identical. 
For this reason, we will restrict our attention to the case where $\hat{H}_{xy}\neq0$. Hence, the invariant $Z_3$ 
is given by
\begin{equation}\label{formofZ3}
	Z_3(t,z,r)=\int\dd t'\dd z'\mathcal{Z}_3^{(xy)}(t-t',z-z',r)\hat{H}_{xy},
\end{equation}
where $\mathcal{Z}_3^{(xy)}$ is the bulk-to-boundary propagator\footnote{The superscript index in the parenthesis specifies the choice of the non-zero sources.}.

Transforming  (\ref{es3}) into the $(w,\rho,r)$-coordinates with $\mathcal{Z}_3^{(xy)}$ given by (\ref{thetheansatz}), we 
find the solution at $\OO(\mu)$,
\begin{equation}
	\mathcal{Z}_3^{(xy)}\Big|_{\mu^1}=\frac{\mu \left(w^6+w^4+6 w^2-2 \rho ^2 \left(w^4+2 w^2+3\right)-12\right)}{5 \pi  r^2 w^8}.
\end{equation}
As expected, there are no log terms in this case. At $\OO(\mu^2)$ we find 
\begin{equation}
	\begin{split}
		\mathcal{Z}_3^{(xy)}\Big|_{\mu^2}=&\frac{\mu^2}{4200 \pi  r^6 w^{10}} \Big[120 w^{10} \left(-4 \rho ^2+5 w^2-6\right) (\log (w)-\log (r))+655 w^8\\
		\phantom{\frac11}&+448 w^6+3136 w^4-12656 w^2+56 \rho ^4 \left(10 w^8+20 w^6+35 w^4+44w^2+36\right)
		\\&-4 \rho ^2 \left(750 w^{10}+40 w^8+345 w^6+476 w^4+448 w^2-2016\right)+8064\Big]\\
		&+\frac{2}{\pi  r^6}\left[\left(1-6 \rho ^2\right) a^{8,1(xy)}_{6,0}+a^{8,1(xy)}_{8,0} \left(w^2-8 \rho ^2\right)\right],
	\end{split}
\end{equation}
where the coefficients $a^{8,1(xy)}_{6,0}$ and $a^{8,1(xy)}_{8,0}$ are not fixed by the near-boundary analysis. 
We also see the presence of log terms which are due to the $xy$-integration and the anomalous dimensions of the double-stress tensors.

\subsubsection{$G_{xy,xy}$}\label{holoprvy}

We now use the holographic dictionary to determine the thermal correlator $G_{xy,xy}$. The action for the scalar invariant $Z_3$ (and $Z_1$ and $Z_2$ below) can be obtained by Fourier transforming and Wick rotating the result obtained in \cite{Kovtun:2005ev}:
\begin{equation}\label{S3ss}
	S_3=\frac{\pi^2C_T}{160}\lim_{r\rightarrow\infty}\int\dd t\dd zr^5\left(1-\frac{\mu}{r^4}\right)\partial_rZ_3(t,z,r)Z_3(t,z,r).
\end{equation}
The invariant $Z_3(t,z,r)$ is fully determined by the bulk-to-boundary propagator $\mathcal{Z}_3^{(xy)}$ via 
eq. \eqref{formofZ3}. 
To compute the action \eqref{S3ss}, we expand $\mathcal{Z}_3^{(xy)}$ near $r=\infty$ as
\begin{equation}
	\mathcal{Z}_3^{(xy)}(t,z,r)=\delta^{(2)}(t,z)+\frac{1}{r^4}\zeta_{3}^{(xy)}(t,z)+\ldots,
\end{equation}
where the dots represent subleading contact terms of $\mathcal{O}(\frac{1}{r^2})$ of the schematic form $\partial^n\delta/r^n$ as well as contributions analytic in $(t,z)$ that are $\mathcal{O}(\frac{1}{r^6})$. As we will see, in the scalar channel $G_{xy,xy}\propto\zeta_3^{(xy)}$.

To proceed, we substitute the bulk-to-boundary propagator into the action (\ref{S3ss}):
\begin{align}
	S_3=\frac{\pi^2C_T}{160}\lim_{r\rightarrow\infty}\!\int&\dd^2x\dd^2x'\dd^2x''(r^5\!-\!\mu r)\partial_r\mathcal{Z}_3^{(xy)}(x\!-\!x',r)\mathcal{Z}_3^{(xy)}(x\!-\!x'',r)\hat{H}_{xy}(x')\hat{H}_{xy}(x'')\nonumber\\
	=&-\frac{\pi^2C_T}{40}\int\dd^2x\dd^2x'\zeta^{(xy)}_3(x-x')\hat{H}_{xy}(x)\hat{H}_{xy}(x')\,,
\end{align}
where in the second line we have integrated the delta function. 
We have  used an abbreviated notation $x=\{t,z\}$, $x'=\{t',z'\}$ and $x''=\{t'',z''\}$ and omitted contact terms
 (see e.g. \cite{Skenderis:2002wp} for a review on holographic renormalization and the treatment of contact terms).

We can now compute the CFT correlator,
\begin{equation}
	G^{bulk}_{xy,xy}=\expval{T_{xy}(t,z)T_{xy}(0,0)}_\beta=-\frac{\delta^2S_3}{\delta\hat{H}_{xy}(t,z)\delta\hat{H}_{xy}(0,0)}=\frac{\pi^2C_T}{20}\zeta^{(xy)}_3(t,z)
\end{equation}
Inserting the explicit bulk solution, we obtain the following results order-by-order in $\mu$:
\begin{align}
	G_{xy,xy}^{(bulk)}\Big|_{\mu^0}=&\frac{\pi C_T}{10(t^2+z^2)^3}\label{eq:Bulkxyxy0}\\
	G_{xy,xy}^{(bulk)}\Big|_{\mu^1}=&\frac{\pi \mu C_T(t^2-z^2)}{100(t^2+z^2)^2}\label{eq:Bulkxyxy1}\\
	G_{xy,xy}^{(bulk)}\Big|_{\mu^2}=&\frac{\pi \mu^2C_T}{4200} \left(3 \left(5 t^2+z^2\right) \log \left(t^2+z^2\right)-\frac{2 \left(75 t^2 z^2+61 z^4\right)}{t^2+z^2}\right)\nonumber\\
	&+\frac{1}{10} \pi  C_T \left(a^{8,1(xy)}_{8,0} \left(t^2-7 z^2\right)-6 z^2 a^{8,1(xy)}_{6,0}\right)\label{eq:Bulkxyxy2}.
\end{align}
We will compare them with the CFT calculations in the next section.

\subsection{Shear channel}

We can repeat the above procedure  to solve the shear channel bulk equation (\ref{es1}) for the sources $\hat{H}_{tx}$ and $\hat{H}_{xz}$
and express the results in terms of  $w$, $\rho$ and $r$.
The explicit expressions are listed in appendix \ref{AppendixB}.
We will now use them to determine $G_{tx,tx}$ and $G_{xz,xz}$ using the AdS/CFT dictionary;
these calculations are summarized below.

\subsubsection{$G_{tx,tx}$ and $G_{xz,xz}$}

The action for the shear channel invariant is given by\footnote{Note the presence of the inverse operator $(\partial_t^2+\partial_z^2)^{-1}$ which is a Fourier transform of $(\omega^2+q^2)^{-1}$  that appears in the action derived in  \cite{Kovtun:2005ev}.} \cite{Kovtun:2005ev}
\begin{equation}\label{shearshearaction}
	\begin{split}
		S_1=&\frac{\pi^2C_T}{160}\lim_{r\rightarrow\infty}\int\dd t\dd z\frac{\left(1-\frac{\mu}{r^4}\right)r^5}{\partial_t^2+\partial_z^2\left(1-\frac{\mu}{r^4}\right)}\partial_rZ_1(t,z,r)Z_1(t,z,r)\\
		=&\frac{\pi^2C_T}{160}\lim_{r\rightarrow\infty}\int\dd t\dd z\left(\frac{r^5}{\partial_t^2+\partial_z^2}+\mathcal{O}(r^2)\right)\partial_rZ_1(t,z,r)Z_1(t,z,r).
	\end{split}
\end{equation}

We begin by turning on  the source $\hat{H}_{tx}$ and follow the same approach as in  subsection \ref{holoprvy}. The shear 
channel invariant is then given by
\begin{equation}\label{formofZ1}
	Z_1(t,z,r)=\int\dd t'\dd z'\mathcal{Z}_1^{(tx)}(t-t',z-z',r)\hat{H}_{tx},
\end{equation}
where $\mathcal{Z}_1^{(tx)}$ is the bulk-to-boundary propagator corresponding to our choice of  source.

The near-boundary expansion of $\mathcal{Z}_1^{(tx)}$ reads
\begin{equation}\label{expansiono}
	\mathcal{Z}_1^{(tx)}=\partial_z\delta^{(2)}(t,z)+\frac{1}{r^4}\zeta_1^{(tx)}+\frac{\log{r}}{r^4}\zeta_{1,log}^{(tx)}+\ldots,
\end{equation}
where the dots correspond to contact terms which are $\mathcal{O}\left(\frac{1}{r^2}\right)$ and non-contact terms which are $\mathcal{O}\left(\frac{1}{r^6}\right)$. Here, however, we encounter $\log r$ terms in the expansion,
\begin{equation}
	\zeta_{1,log}^{(tx)}=-\frac{z\left(840 a^{8,2(tx)}_{8,0}+41 \mu^2\right)}{70 \pi }.
\label{zetazero}
\end{equation}
This term diverges as $r\rightarrow \infty$, unless the value of the coefficient $a^{8,2(tx)}_{8,0}$ is fixed to be
\begin{equation}
	a^{8,2(tx)}_{8,0}=-\frac{41}{840}\mu^2.
\end{equation}

Using the expansion \eqref{expansiono} in the action (\ref{shearshearaction}) and proceeding as in the scalar channel case, we obtain
\begin{equation}
	G_{tx,tx}^{bulk}=\frac{\pi^2C_T}{20}\frac{\partial_z}{\partial_t^2+\partial_z^2}\zeta_1^{(tx)}
\end{equation}
Thus, we arrive at
\begin{align}
	G_{tx,tx}^{(bulk)}\Big|_{\mu^0}=&-\frac{1}{\partial_t^2+\partial_z^2}\frac{3 \pi  C_T \left(t^2-7 z^2\right)}{5 \left(t^2+z^2\right)^5}\label{eq:Bulktxtx0}\\
	G_{tx,tx}^{(bulk)}\Big|_{\mu^1}=&\frac{1}{\partial_t^2+\partial_z^2}\frac{3 \pi  \mu C_T \left(t^4-6 t^2 z^2+z^4\right)}{200 \left(t^2+z^2\right)^4}\label{eq:Bulktxtx1}\\
	G_{tx,tx}^{(bulk)}\Big|_{\mu^2}=&-\frac{1}{\partial_t^2+\partial_z^2}\Bigg[\frac{\pi  \mu^2 C_T}{8400} \left(\frac{2 \left(669 t^4 z^2+804 t^2 z^4+271 z^6\right)}{\left(t^2+z^2\right)^3}+123 \log \left(t^2+z^2\right)\right)\nonumber\\
	&+\frac{3}{5} \pi  a^{8,1(tx)}_{8,0} C_T\Bigg].\label{eq:Bulktxtx2}
\end{align}
Here we keep the inverse operator $(\partial_t^2+\partial_z^2)^{-1}$ explicit, as in the later comparison we will 
act on the corresponding CFT expressions with the operator  $\partial_t^2+\partial_z^2$.
The  correlator $G^{(bulk)}_{xz,xz}$ can be computed in a similar way and the result is presented  order-by-order in $\mu$ in Appendix \ref{AppendixB}.
%If we instead repeat the identical procedure, but with the source $\hat{H}_{xz}$, we find that $a^{8,2(xz)}_{8,0}=\frac{\mu^2}{24}$ in order to cancel the divergent $\log r$ term. The resulting correlator $G^{(bulk)}_{xz,xz}$ is listed order-by-order in $\mu$ in appendix \ref{AppendixB}.

\subsection{Sound channel}

We now consider the sound channel.
Closer inspection reveals that in the sound channel the form of the ansatz must be modified due to a  technical issue present for the diagonal sources. We first explain how it arises and how to treat it and then proceed with the computation of the holographic
TT  correlators.

\subsubsection{Modified ansatz}

We find that for the source $\hat{H}_{tz}$, we are able to extract the corresponding
results in the sound channel using the same ansatz as in the scalar and shear channels. 
However, we observe that if we turn on any of the diagonal sources $\hat{H}_{tt}$, $\hat{H}_{zz}$, $\hat{H}_{xx}$ or $\hat{H}_{yy}$, 
then the ansatz of the form (\ref{thetheansatz}) is no longer valid. 

The reason for this stems from the structure of the vacuum solution $\mathcal{Z}_2^{AdS}$ in these cases. 
Let us take $\hat{H}_{tt}\neq0$ as an example. In this case the AdS propagator 
is  $-(24r^4(w^2-8\rho^2))/(\pi w^{10})$.
From  (\ref{thetheansatz}) it is clear that the ansatz will only be valid if the actual solution of the bulk equations is 
proportional to $(w^2-8\rho^2)$ to all orders in $\mu$. This condition is too restrictive and, as one can show directly, is not satisfied in the case of  (\ref{es2}).

To solve this issue for the diagonal terms, we separate the identity contribution\footnote{
The form of this ansatz is deduced from the structure of the expected CFT results.
We don't explicitly quote the corresponding equations in the paper, but they are the diagonal analogs of 
eqs.  (C.26), (C.29), (C.31) at $\OO(\mu)$ and (C.38), (C.43) and (C.47) at $\OO(\mu^2)$.
}:
\begin{equation}\label{thetheansatzmod}
	\mathcal{Z}_i^{diag}=\mathcal{Z}_i^{AdS}+\left(G^{4,1}_i+G^{4,2}_i\log r\right)+\frac{1}{r^4}\left(G^{8,1}_i+G^{8,2}_i\log r\right)+\ldots,    
\end{equation}
with $G^4$, $G^8$, $\ldots$ defined by
\begin{align}
	&G^{4,j}=\sum_{m=0}^4\sum_{n=-12}^{-4-m}(a^{4,j}_{n,m}+b^{4,j}_{n,m}\log w)w^n\rho^m\\
	&G^{8,j}=\sum_{m=0}^8\sum_{n=-16}^{-m}(a^{8,j}_{n,m}+b^{8,j}_{n,m}\log w)w^n\rho^m\\
	&\phantom{wwwwww}\vdots\nonumber
\end{align}
The upper and lower bounds of the sums were determined in the same way as it was done at the beginning of Section \ref{susucica}.
Ultimately, using the original ansatz (\ref{thetheansatz}) for the off-diagonal sources and the modified one \eqref{thetheansatzmod} for the diagonal ones, allows us  to solve the equation of motion (\ref{es2}). 
The  results are presented in Appendix \ref{AppendixB}.

\subsubsection{$G_{tz,tz}$, $G_{tt,tt}$, $G_{zz,zz}$ and $G_{xx,xx}$}

The action for the sound channel invariant is given by \cite{Kovtun:2005ev}
\begin{equation}
	\begin{split}
		S_2=&-\frac{3\pi^2C_T}{640}\lim_{r\rightarrow\infty}\int\dd t\dd z\frac{r^5\left(1-\frac{\mu}{r^4}\right)}{\left(3\partial_t^2+\partial_z^2\left(3-\frac{\mu}{r^4}\right)\right)^2}\partial_rZ_2(t,z,r)Z_2(t,z,r)\\
		=&-\frac{\pi^2C_T}{1920}\lim_{r\rightarrow\infty}\int\dd t\dd z\left(\frac{r^5}{(\partial_t^2+\partial_z^2)^2}+\mathcal{O}(r^2)\right)\partial_rZ_2(t,z,r)Z_2(t,z,r).
	\end{split}
\end{equation}

Expanding  the bulk-to-boundary propagators for our choices of the sources, eliminating the divergent $\log r$ term and proceeding as above, we eventually obtain

\begin{equation}\label{dufmzeposledna1}
	G^{bulk}_{ab,ab}=\frac{1}{(\partial_t^2+\partial_z^2)^2}D_{ab}\zeta_2^{(ab)},
\end{equation}
where $\zeta_2^{(ab)}$ is the $1/r^4$ term in the near-boundary expansion of the corresponding bulk-to-boundary propagator $\mathcal{Z}_2^{(ab)}$ for the source $\hat{H}_{ab}$ and the operator $D_{ab}$ is given by
\begin{equation}
	{\setlength{\tabcolsep}{11pt}
		\renewcommand{\arraystretch}{1.45}
		\begin{tabular}{c|c|c}\label{table:Dops}
			$a$     &   $b$ &   $D_{ab}$  \\\hline
			$t$     &   $z$ &   $-\frac{\pi^2C_T}{60}\partial_t\partial_z$ \\
			$t$     &   $t$ &   $\frac{\pi^2C_T}{30}\partial_z^2$\\
			$z$     &   $z$ &   $\frac{\pi^2C_T}{30}\partial_t^2$
	\end{tabular}}
\end{equation}

Using the explicit form of the bulk-to-boundary solution we find that the correlation function $G^{(bulk)}_{tz,tz}$ 
is given by
\begin{align}
	G_{tz,tz}^{(bulk)}\Big|_{\mu^0}\!=&-\frac{1}{(\partial_t^2+\partial_z^2)^2}\frac{96 \pi  C_T \left(3 t^4-34 t^2 z^2+3 z^4\right)}{5 \left(t^2+z^2\right)^7}\label{eq:Bulktztz0}\\
	G_{tz,tz}^{(bulk)}\Big|_{\mu^1}\!=&\frac{1}{(\partial_t^2+\partial_z^2)^2}\frac{4 \pi  \mu C_T \left(-t^6+15 t^4 z^2-15 t^2 z^4+z^6\right)}{15 \left(t^2+z^2\right)^6}\label{eq:Bulktztz1}\\
	G_{tz,tz}^{(bulk)}\Big|_{\mu^2}\!=&-\frac{1}{(\partial_t^2+\partial_z^2)^2}\frac{2 \pi  \mu^2 C_T \left(133 t^8-1408 t^6 z^2-110 t^4 z^4+88 t^2 z^6+65 z^8\right)}{1575 \left(t^2+z^2\right)^5},\label{eq:Bulktztz2}
\end{align} 
and analogously for  $G^{(bulk)}_{tt,tt}$ and $G^{(bulk)}_{zz,zz}$ (see Appendix \ref{AppendixB}).

We find that we need to be more careful when analyzing
 the case of $G^{(bulk)}_{xx,xx}$ (and, similarly,  $G^{(bulk)}_{yy,yy}$). 
 If we turn on  the source $\hat{H}_{xx}$ we find a contribution not only from the action $S_2$ but also from $S_3$; the result is
 % In the sound channel this contribution takes the form \eqref{dufmzeposledna1} with $D_{xx}=-\frac{\pi^2C_T}{60}\left(\partial_t^2+\partial_z^2\right)$, while in the scalar channel the sources $\hat{H}_{xx}=-\hat{H}_{yy}\neq0$ lead to the same solution as  for $\hat{H}_{xy}$, which was determined in Section \ref{sekskalal}. Ultimately, we get the boundary correlator as
\begin{equation}\label{dufmzeposledna2}
	G^{bulk}_{xx,xx}=G^{bulk}_{xy,xy}-\frac{\pi^2C_T}{60}\frac{1}{\left(\partial_t^2+\partial_z^2\right)}\zeta_2^{(xx)}.
\end{equation}
The resulting expression for $G^{bulk}_{xx,xx}$ can be found in Appendix \ref{AppendixB}.
In the following section we will compare these results to their CFT counterparts.

\section{Stress tensor thermal two-point functions}\label{CFTside4d}
In this section we study the stress tensor two-point function on $S^1_\beta\times \mathbb{R}^{d-1}$, where $\beta=T^{-1}$ is the inverse temperature, in holographic CFTs, that is, CFTs with large central charge $C_T\gg1$ and a large gap in the spectrum of higher-spin single-trace operators $\Delta_{\rm gap}\gg1$. The case of the purely scalar correlator is reviewed and extended to the integrated correlator in Appendix \ref{AppendixA}, it serves as a useful toy model to study before considering the technically more complicated spinning correlator. Using the stress tensor OPE, we isolate the contribution from multi-stress tensor operators $[T^k]_J$ and read off the CFT data (OPE coefficients, thermal one-point functions and anomalous dimensions) via a comparison to the bulk calculations of metric perturbations around a black hole background in the previous section. In particular, we read off the anomalous dimensions of multi-stress tensor operators of the schematic form $:T_{\mu\nu}T_{\rho\sigma}:$, $:{T_{\mu}}^\rho T_{\rho\nu}:$ and $:T^{\rho\sigma}T_{\rho\sigma}:$ with spin $J=4,2,0$, respectively. We further compute the subleading $\OO(C_T^{-1})$ corrections which determine the near-lightcone behavior of the correlators.

\subsection{OPE expansion and multi-stress tensor contributions}
The contributions of the multi-stress tensor operators to the thermal two-point function of the stress tensor in \eqref{eq:defHTTH} can be computed using the OPE, which can be schematically written as
\begin{equation}\label{eq:TTOPE}\begin{aligned}
		T_{\mu\nu}(x)\times T_{\rho\sigma}(0)\sim \frac{1}{x^{2d}}\Big[&1+x^d\sum_{i=1}^3\lambda^{(i)}_{TTT}A_{\mu\nu\rho\sigma}^{(i),\,\alpha\beta} T_{\alpha\beta}(0)\\
		&+x^{2d}\sum_{J=0,2,4}\sum_{i\in i_J}\lambda^{(i)}_{TT[T^2]_J}B^{(i),\,\mu_1\ldots\mu_J}_{\mu\nu\rho\sigma}[T^2]_{\mu_1\ldots\mu_J}(0)+\ldots\Big],
\end{aligned}\end{equation}
where $[T^k]_{\mu_1\ldots \mu_J}$ are spin-$J$ multi-stress tensor operators, the ellipses denote higher multi-trace operators and their descendants and $i_0=\{1\}$, $i_2=\{1,2\}$ and $i_4=\{1,2,3\}$. On $S^1_\beta \times \mathbb{R}^{d-1}$ only multi-stress tensors $[T^k]_{\mu_1\ldots \mu_J}$ with dimension $\Delta_{k,J}=dk+\OO(C_T^{-1})$ contribute\footnote{In other words, only operators $[T^k]_J$ with no derivatives but various contractions of the indices survive. We therefore denote these operators by the total spin $J$ and the number of stress tensors $k$. Note also that descendants do not contribute to the two-point function on $S^1_\beta\times \mathbb{R}^{d-1}$.}. Here the label $(i)$ denotes the different structures appearing in the OPE of spinning operators. The structures $A_{\mu\nu\rho\sigma}^{(i),\,\alpha\beta}$ and $B^{(i),\,\mu_1\ldots\mu_J}_{\mu\nu\rho\sigma}$ are further fixed by conformal symmetry and depend on $x^{\mu}/|x|$. Upon inserting the OPE \eqref{eq:TTOPE} in the thermal two-point function \eqref{eq:defHTTH}, we find that each term consists of a product of a kinematical piece and the thermal one-point functions $\langle [T^k]_J\rangle_\beta$, weighted by the OPE coefficients $\lambda^{(i)}_{TT[T^k]_J}$. The thermal one-point functions are fixed by symmetry up to an overall coefficient (see e.g.\ \cite{El-Showk:2011yvt,Iliesiu:2018fao})
\begin{equation}\label{eq:thermalOnePt}
	\langle [T^k]_{\mu_1\ldots\mu_J}\rangle_{\beta} = {b_{[T^k]_{J}}\over \beta^{\Delta_{k,J}}}(e_{\mu_1}\cdots e_{\mu_J}-{\rm traces}),
\end{equation}
where $e_\mu$ is a unit vector on $S^1_\beta$. Rather than using the explicit OPE \eqref{eq:TTOPE} together with the thermal expectation value \eqref{eq:thermalOnePt}, we will use the conformal block expansion in a scalar state and take the OPE limit, see Appendix \ref{app:SpinningBlocks} and \cite{Karlsson:2021duj}. To read off the CFT data we compare this to the bulk computations in a planar black hole background given in Section \ref{Section:Bulk}. The bulk result is shown to be consistent with the OPE expansion and we determine the $\OO(C_T^{-1})$ anomalous dimensions $\gamma^{(1)}_J$ of the double-stress tensor operators of the schematic form $:T_{\mu\nu}T_{\rho\sigma}:$, $:{T_{\mu}}^\rho T_{\rho\nu}:$ and $:T^{\rho\sigma}T_{\rho\sigma}:$. We further determine the product of coefficients $\langle [T^2]_{J=0,2,4}\rangle_\beta\lambda^{(i)}_{TT[T^2]_J}$ to leading order in $C_T^{-1}$ and partially at subleading order. In particular, the leading lightcone behavior of the correlators is determined. 

Let us now review the expected scaling with $C_T$ due to multi-stress tensors appearing in the OPE. The central charge $C_T$ is defined by the stress tensor two-point function in the vacuum
\begin{equation}\label{eq:normT}
	\langle T_{\mu\nu}(x)T_{\rho\sigma}(0)\rangle = {C_T\over x^{2d}}\Big[{1\over 2}(I_{\mu\rho}I_{\nu\sigma}+{1\over 2}I_{\mu\sigma}I_{\nu\rho})-{1\over d}\delta_{\mu\nu}\delta_{\rho\sigma}\Big],
\end{equation}
where $I_{\mu\nu}=I_{\mu\nu}(x)=\delta_{\mu\nu}-\frac{2x_\mu x_\nu}{x^2}$.  The CFT data is encoded in a perturbative expansion in $C_T^{-1}$ and a generic $k$-trace operator $[\OO^k]$ with dimension $\Delta_k$ gives the following contribution in the OPE limit\footnote{In general, the OPE expansion is a complicated function of $x^\mu$, below we just keep the scaling with $|x|$. We further suppress the indices of the operators appearing in the OPE.} $|x|/\beta\to0$:
\begin{equation}\label{eq:TkCorr}
	\langle T_{\mu\nu}(x)T_{\rho\sigma}(0)\rangle_\beta|_{[\OO^k]} \propto |x|^{\Delta_{k,J}-2d}\frac{\langle T_{\mu\nu}T_{\rho\sigma}[\OO^k] \rangle\langle[\OO^k]|\rangle_\beta}{\langle [\OO^k][\OO^k]\rangle}.
\end{equation}
Here we are interested in the case of multi-trace stress tensor operators $[\OO^k]=[T^k]_J$ which have a natural normalization
\begin{equation}\label{eq:MultiTNorm}
	\langle [T^k]_J[T^k]_J\rangle \sim C_T^k,
\end{equation}
which follows from the completely factorized contribution. In holographic CFTs dual to semi-classical Einstein gravity, the connected part of correlation functions of stress tensors is proportional to $C_T$:
\begin{equation}\label{eq:CtTTTk}
	\langle T_{\mu\nu}T_{\rho\sigma} [T^{k\neq 2}]_J\rangle \sim C_T.
\end{equation}
An important exception to \eqref{eq:CtTTTk} occurs for $k=2$ where there is a disconnected contribution such that 
\begin{equation}
	\langle T_{\mu\nu}T_{\rho\sigma} [T^2]_J\rangle \sim C_T^2+\ldots,
\end{equation}
where the dots refer to subleading corrections in $C_T^{-1}$ which will play an important role later. Lastly, the expectation value of a multi-stress tensor operator in the thermal state has the following scaling with $C_T$
\begin{equation}\label{eq:MultiTOHScaling}
	\langle[T^k]_J\rangle_\beta \sim {C_T^k\over \beta^{dk}},
\end{equation} 
where we also included the dependence on $\beta$ which is fixed on dimensional grounds.

Using \eqref{eq:MultiTNorm}-\eqref{eq:MultiTOHScaling}, we see that the contribution of multi-stress tensor operators $[T^k]_J$ with dimensions $\Delta_{k,J}=dk+{\cal O}(C_T^{-1})$ to the stress tensor two-point function in the thermal state has the following scaling with $C_T$ for $k\neq 2$
\begin{equation}\label{eq:TkContribution}
	\langle T_{\mu\nu}(x)T_{\rho\sigma}(0)\rangle_\beta|_{[T^{k\neq 2}]_J} \propto {1\over x^{2d}}C_T \left({x\over\beta}\right)^{dk},
\end{equation}

Meanwhile, for $k=2$, the double stress tensor contributions $[T^2]_{J=0,2,4}$ to the thermal two-point function give rise to the disconnected part of the correlator due to the fact that the three-point function $\langle T_{\mu\nu}T_{\rho\sigma} [T^2]_J\rangle \sim C_T^2$, compared to the $\OO(C_T)$ contribution from the connected part. The contribution at $\OO(C_T)$ will therefore contain the first subleading correction to the OPE coefficients $\lambda^{(i)}_{TT[T^2]_J}$, the corrections to the thermal one-point functions, as well as the anomalous dimensions of the double-stress tensor operators.

We define coefficients $\rho_{i,J}$ for the double-stress tensor $[T^2]_J$ with dimensions $\Delta_J:=\Delta_{2,J}$ by: 
\begin{equation}\label{eq:defRho}\begin{aligned}
		\hat{G}_{\mu\nu,\rho\sigma}(x)|_{\mu^2}=|x|^{-8}\Big[\rho_{1,0}g_{\Delta_{0},0,\mu\nu,\rho\sigma}(x)+\sum_{i=1,2}\rho_{i,2} g^{(i)}_{\Delta_{2},2,\mu\nu,\rho\sigma}(x)\\
		+\sum_{i=1,2,3}\rho_{i,4} g^{(i)}_{\Delta_{4},4,\mu\nu,\rho\sigma}(x)\Big],
\end{aligned}\end{equation}
where $\hat{G}_{\mu\nu,\rho\sigma}(x):=\langle T_{\mu\nu}(x)T_{\rho\sigma}(0)\rangle_{\beta }$ is the thermal correlator and $g^{(i)}_{\Delta,J,\mu\nu,\rho\sigma}$ can be obtained by taking the OPE limit of the conformal blocks in the differential basis \cite{Costa:2011mg,Costa:2011dw}, see Appendix \ref{app:SpinningBlocks}. The coefficients $\rho_{i,J}$ are therefore products of OPE coefficients and thermal one-point functions, see \eqref{eq:TkCorr}.  The coefficients $\rho_{i,J}$ and the anomalous dimensions $\gamma_J$ have a perturbative expansion in $C_T^{-1}$
\begin{equation}\begin{aligned}\label{eq:OPEDataTT}
		\rho_{i,J} &= \rho_{i,J}^{(0)}\Big[1+{\rho^{(1)}_{i,J}\over C_T}+\OO(C_T^{-2})\Big],\\
		\Delta_{J} &= 2d+{\gamma^{(1)}_J\over C_T}+\OO(C_T^{-2}),
\end{aligned}\end{equation} 
and lead to the following schematic contribution to the stress tensor two-point function from $[T^2]_J$:\footnote{We stress that \eqref{eq:doubleStressCont} only contains the scaling with $|x|\to0$ while the explicit expression has a more complicated dependence on $x^\mu$ captured in \eqref{eq:defRho}.}
\begin{equation}\label{eq:doubleStressCont}\begin{aligned}
		\hat{G}_{\mu\nu,\rho\sigma}(x)|_{[T^2]_J} &\propto \sum_{i\in i_J}\rho_{i,J}|x|^{\gamma_J^{(1)}}\\
		&\propto \sum_{i\in i_J}\rho_{i,J}^{(0)}\Big[1+{1\over C_T}\Big(\rho^{(1)}_{i,J}+\gamma^{(1)}_J\log|x|\Big)+\OO(C_T^{-2})\Big].
\end{aligned}\end{equation}
Note that the number of structures for the three point functions $\langle T_{\mu\nu} T_{\rho\sigma} [T^2]_{J=0,2,4}\rangle$ is (in $d\geq 4$) $1,2,3$ for $J=0,2,4$, respectively, giving a total of $6$ different structures at this order. From now on we will mainly consider $d=4$.
\subsection{Thermalization of heavy states}
The thermal one-point function of an operator $\OO$ with dimension $\Delta$ and spin $J$ on $S_\beta^1\times \mathbb{R}^{d-1}$ is fixed up to an overall coefficient $b_\OO$ \cite{El-Showk:2011yvt,Iliesiu:2018fao}
\begin{equation}
	\langle \OO_{\mu_1\ldots \mu_J}\rangle_{\beta} = {b_\OO\over \beta^\Delta}(e_{\mu_1}\cdots e_{\mu_J}-{\rm traces}),
\end{equation}
where $e^\mu$  is a unit vector along the thermal circle. To leading order in the $C_T^{-1}$ expansion, we expect multi-stress tensor operators to thermalize in heavy states $|\psi\rangle =|\OH\rangle$ with scaling dimension $\Delta_H\sim C_T$: (see \cite{Karlsson:2021duj} for a discussion on the thermalization of multi-stress tensors and \cite{Lashkari:2016vgj,Lashkari:2017hwq} for a discussion on ETH in CFTs.)
\begin{equation}\label{eq:therm}
	\langle [T^k]_J\rangle_H \approx \langle [T^k]_J\rangle_\beta,
\end{equation}
where we have suppressed the indices. This statement holds to leading order in $C_T^{-1}$. In particular, thermalization of the stress tensor $\langle T_{\mu\nu}\rangle_H=\langle T_{\mu\nu}\rangle_{\beta}$\footnote{In this paper we will take the large volume limit ${\beta\over R}\to0$ of this equation and further set $R=1$.} leads to the following relation\footnote{See e.g. eq. (6.9) of \cite{Karlsson:2021duj}.} between $\beta$ and the scaling dimension $\Delta_H$ in $d=4$
\begin{equation}\label{eq:Temp}
	{b_{T_{\mu\nu}}\over \beta^4} = -{\mu C_T S_4\over 40},
\end{equation}
where $\mu$ is given by\footnote{Note that the definition of $C_T$ differs by a factor of $S_d^2$ compared to \cite{Kulaxizi:2018dxo}.}
\begin{equation}\label{eq:defMu}
	\mu = {4\Gamma(d+2)\over (d-1)^2\Gamma({d\over 2})^2 S_d^2}{\Delta_H\over C_T}
\end{equation}
and $S_{d}={2\pi^{d\over 2}\over \Gamma({d\over 2})}$. 

To leading order in $C_T^{-1}$, the multi-stress tensor operators are expected to thermalize while the expectation value in the heavy state and the thermal state might differ at subleading order. 
%This was already seen in $d=2$ in Section \ref{Section:2d}.
 As evident from \eqref{eq:OPEDataTT}, the $\OO(C_T\mu^2)$ part of the correlator contains corrections to the dynamical data that are subleading in $C_T^{-1}$ . 
%When  comparing these results to the corresponding bulk results computed in the black hole background these are therefore understood as corrections to the thermal one-point functions of these operators. 
More specifically, $\rho_{i,J}^{(1)}$ contain the following terms
\begin{equation}\label{eq:correction}
	\rho_{i,J}^{(1)}= \lambda^{(i,1)}_{TT[T^2]_J}+b_{[T^2]_J}^{(1)},
\end{equation}
where $\lambda^{(i,1)}_{TT[T^2]_J}$ and $b_{[T^2]_J}^{(1)}$ are the subleading $C_T^{-1}$ corrections to the OPE coefficients and the thermal one-point functions, respectively. 
\subsection{Identity contribution}
In this section we compare the contribution of the identity operator in the $T_{\mu\nu}\times T_{\rho\sigma}$ OPE on the CFT side using \eqref{eq:normT} to the bulk results in Section \ref{Section:Bulk}. To make a comparison to the bulk calculation, we integrate \eqref{eq:normT} over the $xy$-plane 
\begin{equation}\label{eq:Identity}\begin{aligned}
		G_{xy,xy}|_{\mu^0} &=\frac{\pi  C_T}{10 \left(t^2+z^2\right)^3},\\
		G_{tx,tx}|_{\mu^0} &=-\frac{\pi  C_T\left(t^2-5 z^2\right)}{40 \left(t^2+z^2\right)^4},\\
		G_{tz,tz}|_{\mu^0} &=-\frac{\pi  C_T \left(5 t^4-38 t^2 z^2+5 z^4\right)}{60 \left(t^2+z^2\right)^5},
\end{aligned}\end{equation}
where $G_{\mu\nu,\rho\sigma}$ is the integrated correlator defined in \eqref{eq:defHTTH}. The result for $G_{xy,xy}$ in \eqref{eq:Identity} agrees with \eqref{eq:Bulkxyxy0} obtained in the bulk. In order to compare the remaining two polarizations $G_{tx,tx}$ and $G_{tz,tz}$, we further apply the differential operator $(\partial_t^2+\partial_z^2)^p$ with $p=1,2$, respectively, to match these CFT results with their bulk counterparts. Doing so, we find that 
\begin{equation}\label{eq:IdentityWDeriv}\begin{aligned}
		(\partial_t^2+\partial_z^2)G_{tx,tx}|_{\mu^0} &=-\frac{3 \pi  C_T\left(t^2-7 z^2\right)}{5 \left(t^2+z^2\right)^5},\\
		(\partial_t^2+\partial_z^2)^2G_{tz,tz}|_{\mu^0} &=-\frac{96 \pi C_T\left(3 t^4-34 t^2 z^2+3 z^4\right)}{5 \left(t^2+z^2\right)^7},
\end{aligned}\end{equation}
which agree with \eqref{eq:Bulktxtx0} and \eqref{eq:Bulktztz0}, respectively.

\subsection{Stress tensor contribution}
In this section we consider the stress tensor contribution. The stress tensor three-point function is fixed up to three coefficients in $d\geq 4$ \cite{Osborn:1993cr}
\begin{equation}
	\langle T_{\mu\nu}(x_1)T_{\rho\sigma}(x_2)T_{\alpha\beta}(x_3)\rangle =\sum_{i=1,2,3}\lambda_{TTT}^{(i)}{\cal I}^{(i)}_{\mu\nu,\rho\sigma,\alpha\beta},
\end{equation}
for three tensor structures ${\cal I}^{(i)}_{\mu\nu,\rho\sigma,\alpha\beta}(x_j)$ determined by conservation and conformal symmetry. One way to parametrize these coefficients is in terms of $(C_T,t_2,t_4)$, for further details and conventions see Appendix \ref{sec:T}. In particular, in holographic CFTs dual to semi-classical Einstein gravity it is known that $t_2=t_4=0$ \cite{Hofman:2008ar}. This fixes two of the coefficients, with the remaining one being fixed by Ward identities in terms of $C_T$ according to \eqref{eq:WardTTT} \cite{Osborn:1993cr}. 

Using the explicit form of the stress tensor conformal block in the OPE limit together with $t_2=t_4=0$, we can find the explicit contribution of the stress tensor to $G_{\mu\nu,\rho\sigma}$, see Appendix \ref{sec:T} for details. To compare to the corresponding bulk results in Section \ref{Section:Bulk} we further need to integrate the correlator over the $xy$-plane. This is done in Appendix \ref{sec:T} and one finds:
\begin{equation}\begin{aligned}\label{eq:TResult}
		&G_{xy,xy}|_{\mu} = {\pi C_T\mu\over 100}{t^2-z^2\over (t^2+z^2)^2},\\
		&G_{tx,tx}|_{\mu} = {\pi C_T\mu\over 800}{-9t^4+6t^2z^2+7z^4\over (t^2+z^2)^3},\\
		&G_{tz,tz}|_{\mu}= {\pi C_T\mu\over 3600}{-105t^6+3t^4z^2+137t^2z^4+77z^6\over (t^2+z^2)^4}. 
\end{aligned}\end{equation}
The result for $G_{xy,xy}$ in \eqref{eq:TResult} agrees with \eqref{eq:Bulkxyxy1}. For the remaining polarizations we apply the relevant differential operators to find
\begin{equation}\label{eq:TtxtxRes}
	(\partial_t^2+\partial_z^2)G_{tx,tx}|_\mu = {3\pi C_T\mu\over 200}{t^4-6t^2z^2+z^4\over (t^2+z^2)^4}
\end{equation}
and 
\begin{equation}\label{eq:Ttztz}
	(\partial_t^2+\partial_z^2)^2G_{tz,tz}|_{\mu} =-{4\pi C_T\mu\over 15}{t^6-15t^4z^2+15t^2z^4-z^6\over(t^2+z^2)^6}.
\end{equation}
Upon comparing \eqref{eq:TtxtxRes} with \eqref{eq:Bulktxtx1} and \eqref{eq:Ttztz} with \eqref{eq:Bulktztz1} we find perfect agreement between the bulk and the CFT calculation.

\subsection{Double stress tensor contributions}
In this section we consider the contribution due to the double-stress tensor operators of the schematic form $:T_{\mu\nu}T_{\rho\sigma}:$, $:{T_{\mu}}^\rho T_{\rho\nu}:$ and $:T^{\rho\sigma}T_{\rho\sigma}:$. These are captured by \eqref{eq:defRho} with $\Delta_J$ and $\rho_{i,j}$ given by \eqref{eq:OPEDataTT}. Details on the conformal blocks are given in Appendix \ref{app:SpinningBlocks}. At $\OO(C_T^2\mu^2)$ we see from \eqref{eq:doubleStressCont} that there are $6$ undetermined coefficients $\rho_{i,J}^{(0)}$ and at $\OO(C_T\mu^2)$ there is a total of $9$ coefficients, in particular, the $6$ coefficients $\rho_{i,J}^{(1)}$ and the $3$ anomalous dimensions $\gamma_{J}^{(1)}$:
\begin{equation}\label{eq:opeDataToBeDet}
	X= \{\rho_{1,0}^{(1)},\rho_{1,2}^{(1)},\rho_{2,2}^{(1)},\rho_{1,4}^{(1)},\rho_{2,4}^{(1)},\rho_{3,4}^{(1)},\gamma^{(1)}_{0},\gamma^{(1)}_{2},\gamma^{(1)}_{4}\}.
\end{equation}

Note that unlike the conformal data discussed so far, which are largely determined by Ward identities, the
results of this Section follow from the dynamics of the five-dimensional Einstein-Hilbert gravity with a negative
cosmological constant.

\subsubsection{Disconnected part}
As expected from thermalization, the $\OO(C_T^2\mu^2)$ disconnected contribution to the stress tensor two-point function  in the thermal states factorizes and is independent of the position $x$:
\begin{equation}
	\hat{G}_{\mu\nu,\rho\sigma} =\langle T_{\mu\nu}\rangle_{\beta} \langle T_{\rho\sigma} \rangle_{\beta}(1+\OO(C_T^{-1})),
\end{equation}
where $\beta$ is the inverse temperature related to $\mu$ by \eqref{eq:Temp}. In particular, only the diagonal terms of $\langle T_{\mu\nu}\rangle_\beta$ are non-zero:
\begin{equation}\begin{aligned}\label{eq:nonDiagMFT}
		\hat{G}_{xy,xy}=0+\OO({C_T\mu^2}),\\
		\hat{G}_{tx,tx}=0+\OO({C_T\mu^2}),\\
		\hat{G}_{tz,tz}=0+\OO({C_T\mu^2}),
	\end{aligned}
\end{equation}
while 
\begin{equation}\begin{aligned}\label{eq:ttttMFT}
		\hat{G}_{tt,tt}= \left({3\over 4}\right)^2{b_{T_{\mu\nu}}^2\over\beta^{8}}\Big[1+\OO(C_T^{-1})\Big].
	\end{aligned}
\end{equation}
Comparing the conformal block expansion in \eqref{eq:defRho} to \eqref{eq:nonDiagMFT}, we find that $5$ out of $6$ of the leading order coefficients $\rho_{i,J}^{(0)}$ are determined in terms of the remaining undetermined coefficient $\rho_{1,0}^{(0)}$:
\begin{equation}\label{eq:MFTsoln}\begin{aligned}
		\rho^{(0)}_{1,2} =&  {324\over 7}\rho^{(0)}_{1,0},\\
		\rho^{(0)}_{2,2} =& {-1728\over 7}\rho^{(0)}_{1,0},\\
		\rho^{(0)}_{1,4} =& {160\over 7}\rho^{(0)}_{1,0},\\
		\rho^{(0)}_{2,4} =& {-1760\over 7}\rho^{(0)}_{1,0},\\
		\rho^{(0)}_{3,4} =& {-480\over 7}\rho^{(0)}_{1,0}.
	\end{aligned}
\end{equation}

The remaining coefficient is fixed by imposing \eqref{eq:ttttMFT} which gives 
\begin{equation}\label{eq:rhothreefour}
	\rho^{(0)}_{1,0} = {\pi^4 \mu^2 C_T^2\over 480000}.
\end{equation}

\subsubsection{Corrections to double stress tensor CFT data}
At $\OO(C_T\mu^2)$ there is a total of $9$ coefficients that fix $G_{\mu\nu,\rho\sigma}$. The goal of this section is to (partially) determine the CFT data (\ref{eq:opeDataToBeDet}) by comparing the conformal block decomposition at $\OO(C_T\mu^2)$ to the bulk calculations in Section \ref{Section:Bulk}. In particular, our analysis will allow us to extract the anomalous dimensions $\gamma_J^{(1)}$ of double-stress tensors $[T^2]_J$, $J=0,2,4$ as well as the near-lightcone behavior of the correlators.

In order to do so we again need to integrate the correlator over the $xy$-plane. This is divergent, as is manifest from dimensional analysis (see also \eqref{eq:TkContribution}). We will tame this divergence by including a factor of $ |x|^{-\epsilon}$ in the integrals which produces simple poles as $\epsilon\to 0$\footnote{Alternatively, one can introduce an IR cutoff in the integrals and the results for the anomalous dimensions and the coefficients $\rho^{(1)}_{i,J}$ will remain the same.}. These will then be absorbed in the undetermined bulk coefficients, see \eqref{eq:solACoeff}.

We will fix the CFT data by comparing the polarizations, $G_{xy,xy}$, $G_{tx,tx}$ and $G_{tz,tz}$, with the corresponding conformal block decomposition given in \eqref{eq:xyxyCFT}, \eqref{eq:txtxCFT} and \eqref{eq:tztzCFT}, with the bulk results given in \eqref{eq:Bulkxyxy2}, \eqref{eq:Bulktxtx2} and \eqref{eq:Bulktztz2}, respectively. For the latter two polarizations, we apply the differential operators $(\partial_t^2+\partial_z^2)^p$, with $p=1,2$, on the OPE expansion in order to match against the bulk calculations, just as for the identity and stress tensor operator, which give
\begin{equation}\label{eq:matching}
	\begin{aligned}
		G_{xy,xy}^{(CFT)}-G_{xy,xy}^{(bulk)}\Big|_{\mu^2 C_T} = 0,\\
		(\partial_t^2+\partial_z^2)\left[G_{tx,tx}^{(CFT)}-G_{tx,tx}^{(bulk)}\right]\Big|_{\mu^2 C_T}= 0,\\
		(\partial_t^2+\partial_z^2)^2\left[G_{tz,tz}^{(CFT)}-G_{tz,tz}^{(bulk)}\right]\Big|_{\mu^2 C_T}= 0.
	\end{aligned}
\end{equation}

There is a common solution which unambiguously fixes the anomalous dimensions to the values:
\begin{equation}\label{eq:solAnomDim}\begin{aligned}
		\gamma_{0}^{(1)} &= -{2480\over 63\pi^4},\\
		\gamma_{2}^{(1)} &=-{4210\over 189\pi^4},\\
		\gamma_{4}^{(1)} &= -{1982\over 35\pi^4},\\
	\end{aligned}
\end{equation}
where we note that the anomalous dimensions in \eqref{eq:solAnomDim} are all negative. Further, we find the following relations among three out of the six coefficients $\rho^{(1)}_{i,J}$
\begin{equation}\label{eq:solCoeff}
	\begin{aligned}
		\rho^{(1)}_{2,2} &=  -{14465\over 1296\pi^4}+\rho_{1,2}^{(1)},\\
		\rho^{(1)}_{2,4} &= {379\over 210\pi^4}+\rho_{1,4}^{(1)},\\
		\rho^{(1)}_{3,4} &= {3083\over 1260\pi^4}+\rho_{1,4}^{(1)},
	\end{aligned}
\end{equation}
while the remaining CFT data $\{\rho_{1,0}^{(1)},\rho_{1,2}^{(1)},\rho_{1,4}^{(1)}\}$ is undetermined and the bulk coefficients are given in \eqref{eq:solACoeff}. We have further checked that this solution is consistent with several other polarizations such as $G_{zx,zx}, G_{tx,zx},G_{zz,zz}$ and $G_{tt,tt}$ by inserting \eqref{eq:solAnomDim}, \eqref{eq:solCoeff} and \eqref{eq:solACoeff} in the OPE expansion and comparing to the explicit bulk calculations. Comparing $G_{xx,xx}$ from the CFT to the bulk calculation, one finds one more linearly independent equation \footnote{The reason for this can be seen from \eqref{dufmzeposledna1} and the table in  (\ref{table:Dops}), when comparing to the CFT result we only apply a differential operator of degree $2$ for the $G_{xx,xx}$ polarization compared to a degree $4$ operator for other polarizations in the sound channel.} \eqref{eq:Axxxx}. The undetermined coefficients $\{\rho_{1,0}^{(1)},\rho_{1,2}^{(1)},\rho_{1,4}^{(1)}\}$ can then be expressed in terms of the undetermined bulk coefficients, see Eqs.\ \eqref{eq:solACoeff} and \eqref{eq:Axxxx}.

\subsection{Lightcone limit}
In this section\footnote{ We thank Kuo-Wei Huang for the discussions which led to the appearance of this section.}   we consider the lightcone limit which is obtained by Wick-rotating $t\to i t$ and taking $v\to 0$, with $u=t-z$ and $v=t+z$. Imposing unitarity on the stress tensor contribution leads to the conformal collider bounds, see e.g.\ 
\cite{Hofman:2008ar,Hofman:2009ug,Kulaxizi:2010jt,Hartman:2015lfa,Li:2015itl,Komargodski:2016gci,Hartman:2016dxc,Hofman:2016awc,Faulkner:2016mzt,Hartman:2016lgu}. Consider now the lightcone limit of the double-stress tensor contribution. One finds the following result for the integrated correlators in the lightcone limit $v\to 0$:
\begin{equation}
	\begin{aligned}
		G_{xy,xy}^{(CFT)}(u,v)|_{\mu^2C_T} &\underset{v\to 0}{=}\pi ^5 \mu ^2 C_T\frac{2 \gamma^{(1)}_4-41\rho^{(1)}_{1,4}+11\rho^{(1)}_{2,4}+30\rho^{(1)}_{3,4}}{48000}{u^3\over v},\cr 
		G_{tx,tx}^{(CFT)}(u,v)|_{\mu^2C_T} &\underset{v\to 0}{=} \pi ^5 \mu ^2 C_T \frac{-113\gamma^{(1)}_4+16(188\rho^{(1)}_{1,4}-77\rho^{(1)}_{2,4}-111\rho^{(1)}_{3,4})}{10752000}{u^4\over v^2},\cr
		G_{tz,tz}^{(CFT)}(u,v)|_{\mu^2C_T} &\underset{v\to 0}{=}\pi ^5 \mu ^2 C_T \frac{ 29\gamma^{(1)}_4-740\rho^{(1)}_{1,4}+308\rho^{(1)}_{2,4}+432\rho^{(1)}_{3,4}}{16128000}{u^5\over v^3},
	\end{aligned}
\end{equation}
where as expected only the spin-4 operator of the schematic form $:T_{\mu\nu}T_{\rho\sigma}:$ contributes\footnote{We have dropped the divergent terms from the integration since these do not contain negative powers of $v$ when $v\to 0$.}. Inserting the solution \eqref{eq:solAnomDim}-\eqref{eq:solCoeff} we find 
\begin{equation}
	\begin{aligned}
		G_{xy,xy}^{(CFT)}(u,v)|_{\mu^2C_T} &\underset{v\to 0}{=}-\frac{\pi  \mu ^2 C_T}{2400}{u^3\over v},\cr 
		G_{tx,tx}^{(CFT)}(u,v)|_{\mu^2C_T} &\underset{v\to 0}{=} -\frac{17 \pi  \mu ^2C_T}{1075200}{u^4\over v^2},\cr
		G_{tz,tz}^{(CFT)}(u,v)|_{\mu^2C_T} &\underset{v\to 0}{=}-\frac{11 \pi  \mu ^2 C_T}{6048000}{u^5\over v^3},
	\end{aligned}
\end{equation}
where we note that the undetermined coefficient $\rho_{1,4}^{(1)}$ drops out in the lightcone limit. The solution in \eqref{eq:solAnomDim}-\eqref{eq:solCoeff} obtained from the bulk computations in Section \ref{Section:Bulk} therefore determines completely the lightcone limit of the correlator to this order.

\section{Discussion}\label{sec:disc}
In this paper we have examined the thermal two-point function of stress tensors in holographic CFTs. 
In the dual picture, this corresponds to studying metric perturbations around a black hole background. 
The thermal two-point function can be decomposed into  contributions of individual operators using the OPE. 
Important contributions to the OPE of two stress tensors include the identity operator, 
 the stress tensor itself, and composite operators made out of the stress tensor  (multi-stress tensors). 
 
The holographic contribution of the identity reproduces the vacuum result.
We also verify that the  stress tensor contribution to the holographic TT correlator agrees with the CFT result,
which is fixed by the three-point functions of the stress tensor in CFTs dual to Einstein gravity
 (our CFT result  agrees with \cite{Kulaxizi:2010jt}).
The leading contribution from the double-stress tensors  corresponds to the disconnected 
part of the correlator.

The  anomalous dimensions and the corrections to the OPE coefficients and thermal one-point functions contribute at next-to-leading order in the $C_T^{-1}$ expansion. 
Comparing the CFT and  holographic calculations, we are able to read off the anomalous dimensions of the double-stress tensors with spin $J=0,2,4$ and obtain partial relations for the subleading corrections to the 
products of OPE coefficients and thermal one-point functions.
It would be interesting to compare our results with the one-loop results of
\cite{Rastelli:2016nze,Alday:2017xua,Aprile:2017bgs,Rastelli:2017udc}.

We are unable to fully determine the double-stress tensor contribution from the near-boundary analysis in the bulk;
indeed some OPE coefficients remain unfixed, although the leading lightcone behavior of the TT correlators at this order
is completely determined.
The situation is reminiscent of the scalar case \cite{Fitzpatrick:2019zqz}, where the contributions of double-trace operators of external scalars were not determined by the near-boundary analysis.
It would be interesting to go beyond the near-boundary expansion to further determine this remaining data. 
In contrast to the scalar case considered in  \cite{Fitzpatrick:2019zqz}, 
in our analysis we further integrated the correlator over a plane to account for different polarizations of the stress tensor. 
This feature introduces some technical complications and it would be interesting to study the correlator without integration. 

Holography provides a powerful tool to study hydrodynamics of strongly coupled quantum field theories and transport coefficients can be read off from the stress tensor two-point function at finite temperature\footnote{The expansion in small momenta compared to the temperature is opposite of the OPE limit and interpolating between the two is challenging. See e.g.\ \cite{Withers:2018srf,Grozdanov:2019kge,Grozdanov:2019uhi,Abbasi:2020ykq,Jansen:2020hfd,Grozdanov:2020koi,Choi:2020tdj,Baggioli:2020loj,Grozdanov:2021gzh} for recent work on the convergence of the hydrodynamic expansion.}. 
The conformal bootstrap provides another window into strongly coupled phenomena when perturbation theory is not applicable. 
While the bootstrap program for vacuum correlators has led to significant developments in the past decade, the corresponding tools for thermal correlators are still developing, see e.g.\ \cite{Caron-Huot:2009ypo,El-Showk:2011yvt,Gobeil:2018fzy,Iliesiu:2018fao,Delacretaz:2018cfk,Delacretaz:2020nit,Alday:2020eua,Karlsson:2021duj,Delacretaz:2021ufg,Dodelson:2022eiz} for related work. In particular, due to an important role played by the stress tensor thermal two-point function, it would be interesting to  better understand the constraints imposed by the conformal bootstrap on this correlator as well as  the implications for a gravitational dual description. 

By the nature of a duality, there are two sides to the same story. In this paper we have used the structure of the stress tensor two-point functions at finite temperature, imposed by conformal symmetry, in order to read off the CFT data by making a comparison
to  the corresponding calculations in the bulk. At the same time, it would be very interesting to study properties of black holes in AdS by bootstrapping thermal correlators on the boundary. We expect a major role to be played by the stress tensor operator and its composites which are related to the metric degrees of freedom in the bulk.

\acknowledgments
This work was supported in part by an Irish Research Council consolidator award.
We thank K-W. Huang, D. Jafferis, M. Kulaxizi, Y-Z. Li, P. O'Donovan, C. Pantelidou for useful discussions and correspondence. Valentina Prilepina gratefully acknowledges support from the Simons Center for Geometry and Physics, Stony Brook University at which some or all of the research for this paper was performed.

\appendix
    \section{Integrated Scalar}\label{AppendixA}

As several new phenomena emerge in the case of integrated correlators, we will first discuss a toy model -- ($d=4$) scalar field, that will serve as a consistency check. We will show that one is able to extract the same OPE data when working with correlators integrated over the $xy$-plane, as in the original approach. 

This appendix is divided into two parts: the first subsection focuses on the case of a scalar field with non-integer scaling dimension, while the second one studies the $\Delta=4$ case, which is more relevant for the stress tensor calculations.

In both subsections we begin by solving the bulk equations of motion where two spatial dimensions are integrated out. We find the solution using the ansatz introduced recently in \cite{Fitzpatrick:2019zqz, Li:2019tpf}, naturally adapted for the integrated case.

On the CFT side we examine the integrated conformal blocks in the OPE limit. In the integer case we explain the emergence of the log term as a result of mixing of the scalar and stress tensor sectors. We also find that further regularization is needed as a result of the integration.

Finally we extract the OPE coefficients\footnote{To the leading order in the large $C_T$ limit.} from the comparison of the bulk calculations and the CFT analysis. 
We conclude that we can extract the same amount of the OPE data in the integrated and non-integrated cases.

\subsection{Scalar field with non-integer scaling dimension}\label{sec:IntegratedScalarNoninteger}

\subsubsection{Bulk-side}
Our aim is to calculate the bulk-to-boundary propagator satisfying the scalar field equation
\begin{align}
	(\Box-m^2)\phi&=0\label{eom1}\\
	\Delta(\Delta-4)-m^2&=0,
\end{align}
on the planar Euclidean AdS-Schwarzschild black hole background
\begin{equation}
	\dd s^2 = r^2(1-\frac{\mu}{r^4})\dd t^2+r^2\dd \vec{x}^2+\frac{1}{r^2(1-\frac{\mu}{r^4})}\dd r^2,\label{metrika}
\end{equation}
where $\vec{x}=(x,y,z)$.

According to the AdS/CFT dictionary we obtain the thermal two-point function as
\begin{equation}
	\expval{\mathcal{O}_L(x_1)\mathcal{O}_L(x_2)}_{\beta}=\lim_{r\rightarrow\infty}r^\Delta\phi(r,x_1,x_2).\label{comp}
\end{equation}
In this subsection we consider the conformal dimension $[\mathcal{O}_L]=\Delta\notin\mathbb{Z}$.

We now integrate over the $xy$-plane, hence we work with the integrated bulk-to-boundary propagator
\begin{equation}
	\Phi(t,z,r)=\iint_{\mathbb{R}^2}\dd x\dd y\,\phi(t,\vec{x},r)\,.
\end{equation}
Equation (\ref{eom1}) in the background (\ref{metrika}) then acquires the form
\begin{equation}
	\left[\Delta(\Delta-4)-r(4+f)\partial_r-r^2f\partial^2_r-\frac{1}{r^2}\partial^2_z-\frac{1}{r^2f}\partial^2_t\right]\,\Phi=0,    
\end{equation}
where $f=1-\frac{\mu}{r^4}$.

To solve this equation, we first transform coordinates $(t,z,r)$ to $(w,\rho,r)$ defined by 
\begin{align}
	\rho&\coloneqq rz\label{defofFPanalogues1}\\
	w^2&\coloneqq 1+r^2t^2+r^2z^2\,.\label{defofFPanalogues2}
\end{align}
These are the natural integrated analogues of the variables introduced in \cite{Fitzpatrick:2019zqz}. In these coordinates we have the following equation for $\Phi$:
\begin{align}
	\big[C_1+C_2\partial_r&+C_3\partial_\rho+C_4\partial_w+C_5\partial^2_r+C_6\partial^2_\rho\nonumber\\
	&+C_7\partial^2_w+C_8\partial_r\partial_\rho+C_9\partial_\rho\partial_w+C_{10}\partial_w\partial_r\big]\,\Phi=0,\label{eom2}
\end{align}
where
\begin{align}
	C_1&=-r^4w^3(\Delta-4)\Delta(r^4-\mu)\\
	C_2&=rw^3(5r^8-6r^4\mu+\mu^2)\\
	C_3&=\rho w^3(5r^8-6r^4\mu+\mu^2)\\
	C_4&=w^2(w^2-1)(5r^8-6r^4\mu+\mu^2)+r^8(1+\rho^2)\nonumber\\
	&\phantom{wwwwwwwwww}+(r^4-\mu)^2(w^2-1)+r^4(r^4-\mu)(w^2-\rho^2)\\
	C_5&=(r^4-\mu)^2r^2w^3\\
	C_6&=(r^4-\mu)^2w^3\rho^2+r^4(r^4-\mu)w^3\\
	C_7&=r^8w(w^2-\rho^2-1)+(r^4-\mu)^2w(w^2-1)^2+r^4(r^4-\mu)w\rho^2\\
	C_8&=2rw^3\rho(r^4-\mu)^2\\
	C_9&=2(r^4-\mu)^2w^2(w^2-1)\rho+2r^4(r^4-\mu)w^2\rho\\
	C_{10}&=2rw^2(r^4-\mu)^2(w^2-1).
\end{align}
Here, using the same logic as in \cite{Fitzpatrick:2019zqz}, we assume the ansatz (focusing only on the solution that corresponds to 
the stress tensor sector on the CFT side, see \cite{Fitzpatrick:2019zqz} for more details) as
\begin{equation}
	\Phi=\Phi_{AdS}\left(1+\frac{G_4}{r^4}+\frac{G_8}{r^8}+\ldots\right),
\end{equation}
where 
\begin{align}
	&G^4=\sum_{m=0}^2\sum_{n=-2}^{4-m}a^4_{n,m}w^n\rho^m\\
	&G^8=\sum_{m=0}^6\sum_{n=-6}^{8-m}a^8_{n,m}w^n\rho^m\\
	&\phantom{wwwwww}\vdots
\end{align}
The vacuum propagator $\Phi_{AdS}$ can be obtained by integrating the known vacuum bulk-to-boundary propagator for the scalar field: 
\begin{equation}
	\Phi_{AdS}(t,z,r)=\iint\dd x\dd y\left[\frac{r}{1+r^2(t^2+x^2+y^2+z^2)}\right]^\Delta =\frac{\pi r^{\Delta-2}}{\Delta-1}\left(1+r^2(t^2+z^2)\right)^{1-\Delta}.
\label{phiads}
\end{equation}
Changing the coordinates in this prefactor to $(w,\rho,r)$ we get
\begin{equation}
	\Phi_{AdS}(w,\rho,r)\propto\frac{r^{\Delta-2}}{w^{2-2\Delta}}.
\end{equation}

Substituting the ansatz into  (\ref{eom2}) we can determine the coefficients $a^j_{n,m}$ as functions of $\Delta$ and $\mu$. In the non-integer case all coefficients $a_{n,m}^4$ and $a_{n,m}^8$ can be found. 
Here we list the nonzero parameters that appear at $\mathcal{O}(\mu^1)$:
\begin{align}
	a^4_{-2,0}&=\frac{2\mu(1-\Delta)}{5}\\
	a^4_{0,0}&=\frac{\mu(\Delta-1)}{5}\\
	a^4_{2,0}&=\frac{3\mu\Delta(\Delta-1)}{20(\Delta-2)}\\
	a^4_{4,0}&=\frac{\mu\Delta(\Delta-1)(3\Delta-10)}{120(\Delta-3)(\Delta-2)}\\
	a^4_{-2,2}&=-\frac{\mu(\Delta-1)}{5}\\
	a^4_{0,2}&=-\frac{\mu\Delta}{10}\\
	a^4_{2,2}&=-\frac{\mu\Delta(\Delta-1)}{30(\Delta-2)}.
\end{align}

\subsubsection{CFT-side}

On the CFT side, the object dual to the scalar field two-point function in the black hole background, is the heavy-heavy-light-light correlator $\expval{\mathcal{O}_H\mathcal{O}_L\mathcal{O}_L\mathcal{O}_H}$.

Decomposing this four-point function into conformal blocks and integrating, we obtain 
\begin{equation}
	\mathcal{G}_{\Delta}\coloneqq\iint\dd x\dd y\expval{\mathcal{O}_H\mathcal{O}_L\mathcal{O}_L\mathcal{O}_H}=\iint\dd x\dd y\sum_{\Delta_i,J}C_{\Delta_i,J}\frac{g_{\Delta_i,J}(Z,\overline{Z})}{(Z\overline{Z})^\Delta}\,,\label{ego}
\end{equation}
where $Z$ and $\overline{Z}$\footnote{We will temporarily use this unusual notation, as we have to distinguish the cross ratios and the space coordinate $z$.} are the cross ratios defined in terms of $t$, $x$, $y$ and $z$ as:
\begin{align}
	Z&=-t-i\sqrt{x^2+y^2+z^2}\\
	\overline{Z}&=-t+i\sqrt{x^2+y^2+z^2},
\end{align}
and $C_{\Delta_i,J}$ is the product of the OPE coefficients corresponding to the primaries with the conformal dimension $\Delta_i$ and spin $J$.

In the heavy-heavy-light-light correlator, the important set of operators contributing in the T-channel  are the multi- stress tensors, which we consider below.
The first nontrivial contribution to the correlator (\ref{ego}) comes from the exchange of the  stress tensor. In the OPE limit the corresponding conformal block is
\begin{equation}
	g_{4,2}(Z,\overline{Z})\approx Z\overline{Z}(Z^2+Z\overline{Z}+\overline{Z}^2).
\end{equation}        

\noindent Hence, at this order we find that  (\ref{ego}) becomes
\begin{equation}\label{CCC1}
	\mathcal{G}_{\Delta}\Big|_{\mu^1}=-C_{4,2}\frac{\pi(t^2+z^2)^{2-\Delta}(t^2(10-3\Delta)+z^2(\Delta-2))}{(\Delta-3)(\Delta-2)}.
\end{equation}

Following the same approach, it is straightforward to obtain the corresponding integrated conformal blocks for the double-trace stress tensors.

\subsubsection{Comparison}

In order to determine the OPE coefficients, we compare the bulk and the CFT results. We connect the two sides by equation (\ref{comp}), which is now of the form
\begin{equation}
	\mathcal{G}_\Delta=\lim_{r\rightarrow\infty}r^\Delta\Phi_{AdS}\left(1+G^T+G^\phi\right).\label{tcomp}
\end{equation}
where $G^T=\frac{G^4}{r^4}+\frac{G^8}{r^8}+\ldots$ 
and $G^\phi$ 
correspond to the stress tensor sector and  the double-trace scalars (possibly dressed with $T_{\mu\nu}$), respectively. As  mentioned above, these two sectors are decoupled for $\Delta\notin\mathbb{Z}$, therefore, we can  consider only the multi-stress tensors. The  stress tensor contribution  to  (\ref{tcomp}) is given by
\begin{equation}\label{jezibaba}
	\mathcal{G}_{\Delta}\Big|_{\mu^1}\!=\lim_{r\rightarrow\infty}\frac{\pi r^{2\Delta-6}}{\Delta-1}\frac{G^4(t,z,r)}{(1+r^2(t^2+z^2))^{\Delta-1}}=\frac{\pi(t^2\!+\!z^2)^{2-\Delta}(t^2(3\Delta-10)\!+\!z^2(2-\Delta))\Delta \mu}{120(\Delta-3)(\Delta-2)},
\end{equation}
where in the second equality we have used the bulk results for $G^4$ .

Comparing (\ref{CCC1}) and \eqref{jezibaba} we extract the OPE coefficient:
\begin{equation}
	C_{4,2}=\frac{\Delta \mu}{120},
\end{equation}
which agrees with  eq.  (3.65) in \cite{Fitzpatrick:2019zqz}.

The OPE coefficients at higher orders in $\mu$ can be determined in a similar way.

\subsection{Scalar field with $\Delta=4$}
\label{sec:IntegratedScalarInteger}

\subsubsection{Bulk-side}

We now consider $\Delta=4$.
The setup for this case is identical to the one above; that is, 
we again need  to solve the bulk equation of motion (\ref{eom2}) but now for $\Delta=4$. 

Here, however, the situation becomes more subtle as some of the OPE coefficients are singular for $\Delta=4$. On the other hand, for integer $\Delta$ the multi-stress tensor sector and double-trace scalar sector are no longer decoupled. We expect the contribution from the $[\mathcal{O}\mathcal{O}]$ to compensate for these divergent parts in the $[T^n]$ OPE coefficients. As a result,  log terms will appear in the solution. We  explain this in more detail in the next subsection. 

In the bulk this leads to a slightly modified ansatz \cite{Li:2019tpf}:
\begin{equation}
	\Phi=\Phi_{AdS}\left(1+\frac{1}{r^4}\left(G^{4,1}+G^{4,2}\log r\right)+\frac{1}{r^8}\left(G^{8,1}+G^{8,2}\log r\right)+\ldots\right),
\end{equation}
where $\Phi_{AdS}$ is the  vacuum propagator (\ref{phiads}) and $G^{4,j}$ and $G^{8,j}$ are given by
\begin{align}
	&G^{4,j}=\sum_{m=0}^2\sum_{n=-2}^{4-m}(a^{4,j}_{n,m}+b^{4,j}_{n,m}\log w)w^n\rho^m\\
	&G^{8,j}=\sum_{m=0}^6\sum_{n=-6}^{8-m}(a^{8,j}_{n,m}+b^{8,j}_{n,m}\log w)w^n\rho^m\\
	&\phantom{wwwwww}\vdots\nonumber
\end{align}
Inserting this ansatz into   (\ref{eom2}), we can determine the coefficients $a_{n,m}^{k,j}$ and $b_{n,m}^{k,j}$.

The result (in the $w$, $\rho$ and $r$ coordinates) is
\begin{equation}\label{sbulkyscalar}
	\begin{split}
		\Phi&=\frac{\pi}{25200r^6w^{10}}\big[8400w^4(r^8+w^6((1-6\rho^2)a^{8,1}_{6,0}+(w^2-8\rho^2)a^{8,1}_{8,0}))\\
		&\phantom{=}+840r^4w^2(-12+6w^2+w^4+w^6-2(3+2w^2+w^4)\rho^2)\mu\\
		&\phantom{=}+(8064-12656w^2+3136w^4+448w^6+655w^8-4(-2016\\
		&\phantom{=}+448w^2+476w^4+345w^6+40w^8+750w^{10})\rho^2\\
		&\phantom{=}+56(36+44w^2+35w^4+20w^6+10w^8)\rho^4\\
		&\phantom{=}+120w^{10}(-6+5w^2-4\rho^2)(\log r+\log w))\mu^2\big]+\mathcal{O}(\mu^3)
	\end{split}
\end{equation}
For  the stress tensor exchange  all log terms vanish and we are also able to determine all the coefficients. 
As expected, we obtain the same results as in the non-integer case.
Meanwhile, we find that for
 the double stress tensor exchange  ($\mu^2$) the coefficients $a^{8,1}_{6,0}$ and $a^{8,1}_{8,0}$ are not fixed by near-boundary analysis.

\subsubsection{CFT-side}\label{sscalcft}

At  $\mathcal{O}(\mu^0)$ and $\mathcal{O}(\mu^1)$ the contribution for $\Delta=4$ will be the same as  for $\Delta\notin\mathbb{Z}$. Let us therefore focus on the $\mu^2$ terms.

Here the double-trace stress tensors mix with the double-trace scalar $[\mathcal{O}\mathcal{O}]$. We thus have to consider four contributions to the correlator at  $\OO(\mu^2)$ -- three from the double stress tensor (we label them by the conformal dimension and the spin: $(\Delta_i,J)$): 
\begin{align}
	T_{\mu\nu}T^{\mu\nu}\quad&\Longleftrightarrow\quad (8,0)\\
	T_{\mu\nu}T^\nu_{\phantom{\mu}\alpha}\quad&\Longleftrightarrow\quad (8,2)\\
	T_{\mu\nu}T_{\alpha\beta}\quad&\Longleftrightarrow\quad (8,4)
\end{align}        
and one contribution from the double-trace scalar:
\begin{equation}
	[\mathcal{O}\mathcal{O}]\quad\Longleftrightarrow\quad(8,0),
\end{equation}
which will mix with the $(8,0)$ contribution from $[T^2]$. This agrees with the fact that it is only the coefficient $C_{8,0}^{\text{\tiny\textit{TT}}}$, that is expected to diverge.

Let us have a closer look at the divergent terms that appear in the corresponding OPE coefficients.
 First, as the coefficient $C_{8,0}^{\text{\tiny\textit{TT}}}$ has a pole in $\Delta=4$ \cite{Fitzpatrick:2019zqz} we can write it as
\begin{equation}
	C_{8,0}^{\text{\tiny\textit{TT}}}=\frac{C_{\text{\tiny sing}}}{\Delta-4}+C^{\text{\tiny\textit{TT}}}_{\text{\tiny reg}},\label{res1}
\end{equation}
where the term $C^{\text{\tiny\textit{TT}}}_{\text{\tiny{reg}}}$ is regular in $\Delta=4$ and $C_{\text{\tiny sing}}$ is the residue. In order to cancel the singular part, the OPE coefficient of the double-trace scalar must also have a pole at $\Delta=4$ with the same residue but with the opposite sign \cite{Fitzpatrick:2019zqz}:
\begin{equation}
	C_{8,0}^{\text{\tiny{$\mathcal{OO}$}}}=-\frac{C_{\text{\tiny{sing}}}}{\Delta-4}+C^{\text{\tiny{$\mathcal{OO}$}}}_{\text{\tiny{reg}}}\label{res2}
\end{equation}
Now, as the conformal block for $J=0$ in the OPE limit is $g_{\Delta',0}\approx(Z\overline{Z})^{\Delta'}$, we can study what happens if the contributions from $[T^2]$ and $[\mathcal{O}\mathcal{O}]$ mix. 
Setting $\Delta=4+\delta$, summing the contributions from $[T^2]$ and $[\mathcal{OO}]$ and then taking the limit $\delta\rightarrow0$ we find
\begin{equation}\label{int}
    \begin{split}
	\mathcal{G}_{4}\Big|_{\mu^2}=\iint\dd x\dd y\Bigg(&C^{\text{\tiny\textit{TT}}}_{\text{\tiny{reg}}}+C^{\text{\tiny{$\mathcal{OO}$}}}_{\text{\tiny{reg}}}-C_{\text{\tiny{sing}}}\log Z\overline{Z}+\\ &C_{8,2}\frac{Z^2+Z\overline{Z}+\overline{Z}^2}{Z\overline{Z}}+C_{8,4}\frac{Z^4+Z^3\overline{Z}+Z^2\overline{Z}^2+Z\overline{Z}^3+\overline{Z}^4}{Z^2\overline{Z}^2}\Bigg).
    \end{split}
\end{equation}
where we used  eqs. (\ref{res1}) and (\ref{res2}).

It is apparent, that this integral is divergent and thus needs to be regulated. In practise we can do this using a
form of the dimensional regularization: we multiply the integrand by a factor $|x|^{-\epsilon}=\left(t^2+x^2+y^2+z^2\right)^{-\frac{\epsilon}{2}}$, integrate and then expand the resulting expression around $\epsilon=0$. In the end, we get
\begin{align}
	\mathcal{G}_{4}\Big|_{\mu^2}=&\frac{8\pi t^2(C_{8,2}-3C_{8,4})}{\epsilon}+\pi\bigg[C_{8,4}\frac{\left(15t^4-2t^2z^2-z^4+12t^2 \left(t^2+z^2\right)\log\left(t^2+z^2\right)\right)}{t^2+z^2}\nonumber\\
	&+C_{8,2}\big(t^2+z^2-4t^2\log\left(t^2+z^2\right)\big)+(t^2+z^2)\Big(C_{\text{\tiny{sing}}}(\log(t^2+z^2)-1)\nonumber\\
	&-C_{\text{\tiny{reg}}}^{\text{\tiny\textit{TT}}}-C_{\text{\tiny{reg}}}^{\text{\tiny{$\mathcal{OO}$}}}\Big)\bigg]+\mathcal{O}(\epsilon^1)\label{cermak}
\end{align}
%In the limit $\epsilon\rightarrow0$ the $\mathcal{O}(\epsilon^1)$ drops out and we are only left with the regular part and the divergent piece that is roughly related to the size of the plane we integrate over.

\subsubsection{Comparison}

In order to compare the bulk and the CFT results, we  apply (\ref{tcomp}). 
Here we are only interested in the double-trace sector
\begin{equation}
	\mathcal{G}_{4}\Big|_{\mu^2}=\lim_{r\rightarrow\infty}r^4\Phi_{AdS}\frac{G^{8,1}+G^{8,2}\log r}{r^8}.\label{comp3}
\end{equation}
The RHS of this relation is obtained by taking the limit of  $\OO(\mu^2)$ term in the bulk result (\ref{sbulkyscalar}), yielding:
\begin{equation}
	\begin{split}
		\mathcal{G}_{4}\Big|_{\mu^2}&=\frac{\pi}{1260}\Bigg[420\Big(\!-6z^2a^{8,1}_{6,0}+(t^2-7z^2)a^{8,1}_{8,0}\Big)\\
		&+\mu^2\bigg(-\frac{2(75t^2z^2+61z^4)}{t^2+z^2}+3(5t^2+z^2)\log\left(t^2+z^2\right)\bigg)\Bigg].\label{RHSf}
	\end{split}
\end{equation}

Comparing (\ref{cermak}) and (\ref{RHSf}) we can extract the coefficients $C_{8,2}$, $C_{8,4}$ and $C_{\text{\tiny{sing}}}$:
\begin{align}
	C_{8,2}&=\frac{\mu^2}{560}\label{val1}\\
	C_{8,4}&=\frac{\mu^2}{720}\label{val2}\\
	C_{\text{\tiny{sing}}}&=\frac{\mu^2}{420},\label{val3}
\end{align}
while for the coefficients $C_{\text{\tiny{reg}}}^{TT}$, $C_{\text{\tiny{reg}}}^{\text{\tiny{$\mathcal{OO}$}}}$ and the parameter $\epsilon$ we get the following relations
\begin{align}
	C_{\text{\tiny{reg}}}^{\text{\tiny{$\mathcal{OO}$}}}+C_{\text{\tiny{reg}}}^{\text{\tiny\textit{TT}}}&=2 a^{8,1}_{6,0}+\frac{7}{3} a^{8,1}_{8,0}+\frac{239 \mu ^2}{2520}\label{sht1}\\
	\frac{1}{\epsilon}&=-\frac{420 (3 a^{8,1}_{6,0}+4 a^{8,1}_{8,0})+47 \mu ^2}{12 \mu ^2}.\label{sht2}
\end{align}

To conclude, our double-trace results for $C_{8,2}$, $C_{8,4}$ and the residual part of $C_{8,0}^{\text{\tiny\textit{TT}}}$ are in perfect agreement with the results for the non-integer case extrapolated to $\Delta=4$, see \cite{Fitzpatrick:2019zqz}, while the remaining CFT data is related to the undetermined coefficients on the bulk side by  eqs (\ref{sht1})-(\ref{sht2}).

Using the same approach in the non-integrated $\Delta=4$ case, one gets the relations \eqref{val1}-\eqref{val3} for $C_{8,2}$, $C_{8,4}$ and $C_{\text{\tiny{sing}}}$, while $C_{\text{\tiny{reg}}}^{\text{\tiny{$\mathcal{OO}$}}}+C_{\text{\tiny{reg}}}^{\text{\tiny\textit{TT}}}$ is related to a single undetermined bulk coefficient by a relation analogous to \eqref{sht1}.

\section{List of bulk results for $Z_1$ and $Z_2$}\label{AppendixB}

In this appendix we list some expressions for the invariants  in the shear and sound channels.

\subsection{Results in the shear channel}

For the source $\hat{H}_{tx}$ we find the following solution of eq. (\ref{es1}) at $\mathcal{O}(\mu^1)$:
\begin{equation}
	\mathcal{Z}_1^{(tx)}\Big|_{\mu^1}=\frac{\mu \rho  \left(96 \left(\rho ^2+2\right)+3 w^6+\left(6-4 \rho ^2\right) w^4-12 \left(\rho ^2+8\right) w^2\right)}{10 \pi  r w^{10}}     
\end{equation}
and at $\mathcal{O}(\mu^2)$:
\begin{equation}
	\begin{split}
		\mathcal{Z}_1^{(tx)}\Big|_{\mu^2}=&\frac{\mu^2 \rho}{8400 \pi  r^5 w^{12}} \Big[-40320 \left(\rho ^2+2\right)^2-4920 w^{12} \log (w)+\left(6920-7280 \rho ^2\right) w^{10}\\
		&+5 \left(272 \rho ^4-2880 \rho ^2+271\right) w^8+40 \left(136 \rho ^4-331 \rho
		^2-154\right) w^6\\
		&+280 \left(33 \rho ^4+26 \rho ^2-268\right) w^4+896 \left(\rho ^4+140 \rho ^2+262\right) w^2\Big]\\
		&-\frac{12 \rho  \left(a^{8,2(tx)}_{8,0} \log (r)+a^{8,1(tx)}_{8,0}\right)}{\pi  r^5},       
	\end{split}
\end{equation}
where $a^{8,1(tx)}_{8,0}$ and $a^{8,2(tx)}_{8,0}$ are undetermined coefficients.

Choosing a source $\hat{H}_{xz}$, we get the bulk result:
\begin{equation}\label{propxz1}
	\mathcal{Z}_1^{(xz)}\Big|_{\mu^1}=-\frac{f_0 \sqrt{-\rho ^2+w^2-1} \left(96 \left(\rho ^2+2\right)+w^6+\left(2-4 \rho ^2\right) w^4-12 \left(\rho ^2+6\right) w^2\right)}{10 \pi  r w^{10}}
\end{equation}
and
\begin{equation}\label{propxz2}
	\begin{split}
		\mathcal{Z}_1^{(xz)}\Big|_{\mu^2}=&\frac{\mu^2 \sqrt{-\rho ^2+w^2-1}}{8400 \pi  r^5 w^{12}} \Big[40320 \left(\rho ^2+2\right)^2-4200 w^{12} \log (w)+120 \left(38 \rho ^2+17\right) w^{10}+\\
		&+5 \left(-272 \rho ^4+1792 \rho ^2+437\right) w^8+8 \left(-680 \rho
		^4+885 \rho ^2+448\right) w^6-\\
		&-168 \left(55 \rho ^4+46 \rho ^2-284\right) w^4-896 \left(\rho ^4+122 \rho ^2+226\right) w^2\Big]+\\
		&+\frac{12 \sqrt{-\rho ^2+w^2-1} \left(a^{8,2(xz)}_{8,0} \log (r)+a^{8,1(xz)}_{8,0}\right)}{\pi  r^5}\,.
	\end{split}
\end{equation}

Using the results for the bulk-to-boundary propagator $\mathcal{Z}_1^{(xz)}$ \eqref{propxz1} and \eqref{propxz2} we obtain the correlator $G_{xz,xz}^{(bulk)}$:

\begin{align}
	G_{xz,xz}^{(bulk)}\Big|_{\mu^0}\!=&-\frac{1}{\partial_t^2+\partial_z^2}\frac{3 \pi  C_T \left(z^2-7 t^2\right)}{5 \left(t^2+z^2\right)^5}\label{GxzxzBulk0}\\
	G_{xz,xz}^{(bulk)}\Big|_{\mu^1}\!=&-\frac{1}{\partial_t^2+\partial_z^2}\frac{3 \pi  \mu C_T \left(t^4-6 t^2 z^2+z^4\right)}{200 \left(t^2+z^2\right)^4}\label{GxzxzBulk1}\\
	G_{xz,xz}^{(bulk)}\Big|_{\mu^2}\!=&\frac{1}{\partial_t^2+\partial_z^2}\Bigg[\frac{\pi  \mu^2 C_T}{8400 \left(t^2+z^2\right)^3} \Big(210 t^6+648 t^4 z^2+6 t^2 z^4-160 z^6\nonumber\\
	&+105 \left(t^2+z^2\right)^3 \log \left(t^2+z^2\right)\Big)-\frac{3}{5} \pi  a^{8,1(xz)}_{8,0} C_T\Bigg]\label{GxzxzBulk2}.
\end{align}

\subsection{Results in the sound channel}

First we list the solutions of the sound channel equations of motion (\ref{es2}) for various polarizations.
For the source $\hat{H}_{tz}$ we get
\begin{equation}
	\mathcal{Z}_2^{(tz)}\Big|_{\mu^1}=\frac{16 \mu \rho  \sqrt{-\rho ^2+w^2-1}}{5 \pi  w^{12}} \Big[-w^6-3 w^4-96 w^2+2 \rho ^2 \left(w^4+4 w^2+60\right)+240\Big]
\end{equation}
and
\begin{equation}
	\begin{split}
		\mathcal{Z}_2^{(tz)}\Big|_{\mu^2}=&-\frac{4 \mu^2 \rho  \sqrt{-\rho ^2+w^2-1}}{315 \pi  r^4 w^{14}} \Big[18144 \left(\rho ^2+2\right)^2+798 w^{12}+\left(1356-584 \rho ^2\right) w^{10}\\
		&+\left(176 \rho ^4-2240 \rho ^2+1779\right) w^8+12
		\left(88 \rho ^4-384 \rho ^2+147\right) w^6\\
		&+336 \left(9 \rho ^4-17 \rho ^2+58\right) w^4+672 \left(7 \rho ^4-55 \rho ^2-131\right) w^2\Big].
	\end{split}
\end{equation}
For the source $\hat{H}_{tt}$ we get
\begin{equation}
	\begin{split}
		\mathcal{Z}_2^{(tt)}\Big|_{\mu^1}=&-\frac{2 \mu}{5 \pi  w^{12}} \Big[8 \rho ^4 \left(w^4+4 w^2+60\right)-8 \rho ^2 \left(w^6+3 w^4+66 w^2-120\right)\\
		&+w^2 \left(w^6+2 w^4+48 w^2-96\right)\Big]
	\end{split}
\end{equation}
and
\begin{equation}
	\begin{split}
		\mathcal{Z}_2^{(tt)}\Big|_{\mu^2}=&\frac{\mu^2}{3150 \pi  r^4 w^{14}} \Big[ 362880 \rho ^2 \left(\rho ^2+2\right)^2+15960 w^{14} \log (w)+120 \left(279 \rho ^2-113\right) w^{12}\\
		&-15 \left(1072 \rho ^4-4048 \rho ^2+593\right) w^{10}+20
		\left(176 \rho ^6-3120 \rho ^4+4083 \rho ^2-294\right) w^8\\
		&+120 \left(176 \rho ^6-1083 \rho ^4+651 \rho ^2-406\right) w^6+1344 (5 \left(9 \rho ^4-24 \rho ^2+91\right) \rho
		^2\\
		&+131) w^4+3360 \left(28 \rho ^6-265 \rho ^4-632 \rho ^2-36\right) w^2\Big]+\frac{a_{0,0}^{8,1(tt)}+a_{0,0}^{8,2(tt)} \log (r)}{r^4}
		.
	\end{split}
\end{equation}
For the source $\hat{H}_{zz}$ we get
\begin{equation}
	\begin{split}
		\mathcal{Z}_2^{(zz)}\Big|_{\mu^1}=&\frac{2 \mu}{5 \pi  w^{12}} \Big[480 \left(\rho ^4+3 \rho ^2+2\right)+w^8+\left(2-8 \rho ^2\right) w^6\\
		&+8 \left(\rho ^4-3 \rho ^2+31\right) w^4+16 \left(2 \rho ^4-43 \rho ^2-72\right)
		w^2\Big]
	\end{split}
\end{equation}
and
\begin{equation}
	\begin{split}
		\mathcal{Z}_2^{(zz)}\Big|_{\mu^2}=&\frac{\mu^2}{630 \pi  r^4 w^{14}} \Big[-72576 \left(\rho ^2+1\right) \left(\rho ^2+2\right)^2+1560 w^{14} \log (w)-24 \left(99 \rho ^2+16\right) w^{12}\\
		&+3 \left(720 \rho ^4-1776 \rho ^2-37\right)
		w^{10}-4 \left(176 \rho ^6-2240 \rho ^4+1791 \rho ^2+75\right) w^8\\
		&-24 \left(176 \rho ^6-781 \rho ^4+177 \rho ^2-1806\right) w^6-4032 \left(3 \rho ^6-5 \rho ^4+37 \rho ^2+81\right)
		w^4\\
		&-672 \left(28 \rho ^6-255 \rho ^4-1032 \rho ^2-848\right) w^2\Big]+\frac{a_{0,0}^{8,1(zz)}+a_{0,0}^{8,2(zz)}(0,0) \log (r)}{r^4}.
	\end{split}
\end{equation}
For the source $\hat{H}_{xx}$ we get
\begin{equation}
	\mathcal{Z}_2^{(xx)}\Big|_{\mu^1}=-\frac{8 \mu}{5 \pi  w^{12}} \Big[60 \left(\rho ^2+2\right)+25 w^4-4 \left(5 \rho ^2+33\right) w^2\Big]
\end{equation}
and
\begin{equation}
	\begin{split}
		\mathcal{Z}^{(xx)}_2\Big|_{\mu^2}=&\frac{\mu^2}{3150 \pi  r^4 w^{14}} \Big[181440 \left(\rho ^2+2\right)^2-11880 w^{14} \log (w)+180 \left(43-60 \rho ^2\right) w^{12}\\
		&+15 \left(176 \rho ^4-1136 \rho ^2+315\right) w^{10}+10 \left(880 \rho
        ^4-2292 \rho ^2+369\right) w^8\\
        &+120 \left(151 \rho ^4-237 \rho ^2-700\right) w^6+672 \left(45 \rho ^4+100 \rho ^2+1084\right) w^4\\
        &+3360 \left(5 \rho ^2 \left(\rho
        ^2-40\right)-406\right) w^2\Big]+\frac{a_{0,0}^{8,1(xx)}+a_{0,0}^{8,2(xx)} \log (r)}{r^4}.
	\end{split}
\end{equation}

Using the prescription (\ref{dufmzeposledna1}) for the sound channel and the solutions above, 
we find  that the correlator order-by-order in $\mu$ for the source $\hat{H}_{tt}$ is given by        
\begin{align}
	G_{tt,tt}^{(bulk)}\Big|_{\mu^0}\!=&\frac{1}{(\partial_t^2+\partial_z^2)^2}\frac{96 \pi  C_T \left(t^4-18 t^2 z^2+21 z^4\right)}{5 \left(t^2+z^2\right)^7}\label{GttttBulk0}\\
	G_{tt,tt}^{(bulk)}\Big|_{\mu^1}\!=&\frac{1}{(\partial_t^2+\partial_z^2)^2}\frac{4 \pi  \mu C_T \left(t^6-15 t^4 z^2+15 t^2 z^4-z^6\right)}{15 \left(t^2+z^2\right)^6}\label{GttttBulk1}\\
	G_{tt,tt}^{(bulk)}\Big|_{\mu^2}\!=&-\frac{1}{(\partial_t^2+\partial_z^2)^2}\frac{2 \pi  \mu^2 C_T \left(-691 t^8+1900 t^6 z^2+1910 t^4 z^4+860 t^2 z^6+133 z^8\right)}{1575 \left(t^2+z^2\right)^5},\label{GttttBulk2}
\end{align}
and for the source $\hat{H}_{zz}$ as
\begin{align}
	G_{zz,zz}^{(bulk)}\Big|_{\mu^0}\!=&\frac{1}{(\partial_t^2+\partial_z^2)^2}\frac{96 \pi  C_T \left(21 t^4-18 t^2 z^2+z^4\right)}{5 \left(t^2+z^2\right)^7}\label{GzzzzBulk0}\\
	G_{zz,zz}^{(bulk)}\Big|_{\mu^1}\!=&\frac{1}{(\partial_t^2+\partial_z^2)^2}\frac{4 \pi  \mu C_T \left(t^6-15 t^4 z^2+15 t^2 z^4-z^6\right)}{15 \left(t^2+z^2\right)^6}\label{GzzzzBulk1}\\
	G_{zz,zz}^{(bulk)}\Big|_{\mu^2}\!=&\frac{1}{(\partial_t^2+\partial_z^2)^2}\frac{2 \pi  \mu^2 C_T \left(-65 t^8-724 t^6 z^2+810 t^4 z^4+140 t^2 z^6+79 z^8\right)}{1575 \left(t^2+z^2\right)^5}.\label{GzzzzBulk2}
\end{align}

Finally, using the relation (\ref{dufmzeposledna2}) we get the $G^{bulk}_{xx,xx}$ in the form
\begin{align}
	G_{xx,xx}^{(bulk)}\Big|_{\mu^0}\!=&\frac{1}{\partial_t^2+\partial_z^2}\frac{24 \pi  C_T}{5 \left(t^2+z^2\right)^4}\label{GxxxxBulk0}\\
	G_{xx,xx}^{(bulk)}\Big|_{\mu^1}\!=&\,0\label{GxxxxBulk1}\\
	G_{xx,xx}^{(bulk)}\Big|_{\mu^2}\!=&\frac{1}{\partial_t^2+\partial_z^2}\Bigg[\frac{\pi  \mu^2 C_T}{3150} \left(126 \log \left(t^2+z^2\right)+\frac{-135 t^4+90 t^2 z^2-71 z^4}{\left(t^2+z^2\right)^2}\right)\nonumber\\
	&-\frac{1}{60} \pi  C_T \left(72 \left({a}_{6,0}^{8,1(xy)}+{a}_{8,0}^{8,1(xy)}\right)+\pi  a_{0,0}^{8,1(xx)}\right)\Bigg],\label{GxxxxBulk2}
\end{align}
where the undetermined coefficients ${a}_{6,0}^{8,1(xy)}$ and ${a}_{8,0}^{8,1(xy)}$ come from the scalar channel contribution and $a_{0,0}^{8,1(xx)}$ from  the sound channel.

\section{Conventions and details on spinning conformal correlators}\label{app:SpinningBlocks}
In this appendix we summarize our conventions and provide some details on spinning conformal correlators in the embedding space that are used in the main part of this paper following \cite{Costa:2011dw,Costa:2011mg}. The basic building blocks are 
\begin{equation}\label{eq:EmbeddSpaceHV}\begin{aligned}
		V_{i,jk}&= {(Z_i\cdot P_j)(P_i\cdot P_k)-(Z_i\cdot P_k)(P_i\cdot P_j)\over P_j\cdot P_k},\\
		H_{ij}&= -2[(Z_i\cdot Z_j)(P_i\cdot P_j)-(Z_i\cdot P_j)(Z_j\cdot P_i)],
\end{aligned}\end{equation}
where $V_1\equiv V_{1,23}$, $V_2\equiv V_{2,31}$ and $V_3\equiv V_{3,12}$. Here $P_i$ and $Z_i$ are null vectors in $\mathbb{R}^{1,d+1}$.

One possible basis for the three-point function of two stress tensors and a spin-$J$ operator with dimension $\Delta$ is given by $(P_{ij}=-2P_i\cdot P_j)$
\begin{equation}\label{eq:GenthreePt}
	\langle T(P_1,Z_1)T(P_2,Z_2)\OO(P_3,Z_3)\rangle = {\sum_{p=1}^{10} x^{(TT\OO)}_{p} Q_p\over  (P_{12})^{d+2-{\Delta+J\over 2}}(P_{23})^{\Delta+J\over 2}(P_{31})^{\Delta+J\over 2}},
\end{equation}
where 
\begin{equation}\label{eq:defQ}\begin{aligned}
		&Q_1 = V_1^2V_2^2V_3^J,\\
		&Q_2 = (H_{23}V_1^2V_2+H_{13}V_2^2V_1)V_3^{J-1},\\
		&Q_3 = H_{12}V_1V_2V_3^J,\\
		&Q_4 = (H_{13}V_2+H_{23}V_1)H_{12}V_3^{J-1},\\
		&Q_5 = H_{13}H_{23}V_1V_2V_3^{J-2},\\
		&Q_6 = H_{12}^2V_3^J,\\
		&Q_7 = (H_{13}^2V_2^2+H_{23}^2V_1^2)V_3^{J-2},\\
		&Q_8 = H_{12}H_{13}H_{23}V_3^{J-2},\\
		&Q_9 = (H_{13}H_{23}^2V_1+H_{23}H_{13}^2V_2)V_3^{J-3},\\
		&Q_{10} = H_{13}^2H_{23}^2V_3^{J-4}.
\end{aligned}\end{equation}
Conservation of the stress tensor further reduces the number of independent structures. In particular, when $\OO=T$ there are $3$ independent structures while for non-conserved operators of dimension $\Delta$ and spin $J=0,2,4$, there are $1,2$ and $3$ independent structures, respectively. However, we will mainly consider the differential basis introduced in \cite{Costa:2011dw} since this is useful when considering the four-point conformal blocks. It is based on multiplication by $H_{12}$ as well as the differential operators
\begin{equation}
	\begin{aligned}
		D_{11} =& (P_1\cdot P_2)(Z_1\cdot {\partial\over \partial P_2})-(Z_1\cdot P_2)(P_1\cdot {\partial\over \partial P_2})\cr
		&-(Z_1\cdot Z_2)(P_1\cdot {\partial\over \partial Z_2})+(P_1\cdot Z_2)(Z_1\cdot {\partial\over \partial Z_2}),\cr
		D_{12} =& (P_1\cdot P_2)(Z_1\cdot {\partial\over \partial P_1})-(Z_1\cdot P_2)(P_1\cdot {\partial\over \partial P_1})+(Z_1\cdot P_2)(Z_1\cdot {\partial\over \partial Z_1}),
	\end{aligned}
\end{equation}
and $D_{22}$ and $D_{21}$ obtained from $D_{11}$ and $D_{12}$ by $1\leftrightarrow 2$. We further define the following differential operators:
\begin{equation}\label{eq:DiffOp}
	\begin{aligned}
		D_1 &= D_{11}^2D_{22}^2\Sigma_L^{2,2},\\
		D_2 &= H_{12}D_{11}D_{22}\Sigma_L^{2,2},\\
		D_3 &= D_{21}D_{11}^2D_{22}\Sigma_L^{3,1}+D_{12}D_{22}^2D_{11}\Sigma_L^{1,3},\\
		D_4 &= H_{12}(D_{21}D_{11}\Sigma_L^{3,1}+D_{12}D_{22}\Sigma_L^{1,3}),\\
		D_5 &= D_{12}D_{21}D_{11}D_{22}\Sigma_L^{2,2},\\
		D_6 &= H_{12}^2\Sigma_L^{2,2},\\
		D_7 &= D_{21}^2D_{11}^2\Sigma_L^{4,0}+D_{12}^2D_{22}^2\Sigma_L^{0,4},\\
		D_8 &= H_{12}D_{12}D_{21}\Sigma_L^{2,2},\\
		D_9 &= D_{12}^2D_{21}^2\Sigma_L^{2,2},\\
		D_{10} &= D_{12}D_{21}^2D_{11}\Sigma_L^{3,1}+D_{21}D_{12}^2D_{22}\Sigma_L^{1,3},
	\end{aligned}
\end{equation}
where $\Sigma_{L}^{m,n}$ denotes the shifts $\Delta_{1}\to\Delta_1+m$ and $\Delta_{2}\to\Delta_2+n$. The three-point functions in the differential basis are then given by 
\begin{equation}\begin{aligned}\label{eq:SpinningThreePt}
		\langle T(P_1,Z_1)&T(P_2,Z_2)\OO_{\Delta,J}(P_3,Z_3)\rangle \cr
		&= \sum_{i=1}^{10}\lambda_{TT\OO_{\Delta,J}}^{(i)}D_i {V_3^J\over P_{12}^{\Delta_1+\Delta_2-\Delta-J}P_{23}^{\Delta+\Delta_2-\Delta_1+J}P_{13}^{\Delta+\Delta_1-\Delta_2+J}},
\end{aligned}\end{equation}
where we kept $\Delta_{1,2}$ to keep track of the action of $\Sigma_L^{m,n}$ in \eqref{eq:DiffOp}.

The spinning conformal partial waves can be obtained from the scalar partial waves $W_{\OO}$: 
\begin{equation}\label{eq:PartialWave}
	W_{\OO} = \left({P_{24}\over P_{14}}\right)^{\Delta_{12}\over 2} \left({P_{14}\over P_{13}}\right)^{\Delta_{34}\over 2}{g_{\Delta,J}^{(\Delta_{12},\Delta_{34})}(u,v)\over P_{12}^{\Delta_1+\Delta_2\over 2}P_{34}^{\Delta_3+\Delta_4\over 2}}
\end{equation}
with $\Delta_{ij}=\Delta_i-\Delta_j$ and the cross-ratios $(u,v)$ are given by 
\begin{equation}\begin{aligned}
		u&= {P_{12}P_{34}\over P_{13}P_{24}},\\
		v&={P_{14}P_{23}\over P_{13}P_{24}}.
\end{aligned}\end{equation}
The scalar conformal blocks are normalized as follows in the limit $u\to 0, v\to 1$:
\begin{equation}
	g_{\Delta,J}^{(\Delta_{12},\Delta_{34})}(u,v) \underset{u\to 0,v\to 1}{\sim} {J!\over (-2)^J({d\over 2}-1)_J}u^{\Delta\over 2}C_J^{({d\over 2}-1)}\Big({v-1\over 2\sqrt{u}}\Big),
\end{equation}
where $C_{J}^{({d\over 2}-1)}$ are Gegenbauer polynomials and $(a)_J$ denotes the Pochammer symbol. The spinning conformal partial waves are then obtained via
\begin{equation}\label{eq:diffRep}
	W_{\OO}^{\{i\}} = {\cal D}_{L}{\cal D}_{R} W_{\OO},
\end{equation}
where 
\begin{equation}\label{eq:DLeft}
	{\cal D}_{L} = H_{12}^{n_{12}}D_{12}^{n_{10}}D_{21}^{n_{20}}D_{11}^{m_1}D_{22}^{m_2}\Sigma_L^{m_1+n_{20}+n_{12},m_2+n_{10}+n_{12}},
\end{equation}
where $i$ labels the structure in the scalar partial wave and ${\cal D}_{R}$ is similarly defined with $1\to3$ and $2\to 4$. The integers $n_{ij}\geq 0$ and $m_{i}$ that labels the structure are determined by the solutions to the following equations ensuring the correct homogeneity under $P\to\alpha P$ and $Z\to\beta Z$:
\begin{equation}
	\begin{aligned}
		m_1 &= J_1-n_{12}-n_{10}\geq 0,\cr
		m_2 &= J_2-n_{12}-n_{20}\geq 0,\cr
		m_0 &= J_0-n_{10}-n_{20}\geq 0,
	\end{aligned}
\end{equation}
where $J=J_0$ is the spin of the exchanged operator. In the case of two spin-$2$ operators at $P_1$ and $P_2$ and scalar operators at $P_3$ and $P_4$, the possible combinations appearing in \eqref{eq:DLeft} can be taken to be the ones given in \eqref{eq:DiffOp}.

We are interested in the OPE limit of the contribution of individual blocks to
\begin{equation}
	\hat{G}(P_i,Z_i):=P_{34}^{\Delta_H}\langle  T(P_1,Z_1)T(P_2,Z_2)\OH(P_3)\OH(P_4)\rangle,
\end{equation}
where $\OH$ is a scalar operator with dimension $\Delta_H$. Using \eqref{eq:diffRep}, we find in this case
\begin{equation}\label{eq:SpinningBlocks}
	\begin{aligned}
		\hat{G}(P_i,Z_i)|&_{\OO_{\Delta,J}} = \sum_{i=1}^{10}\lambda_{TT\OO_{\Delta,J}}^{(i)}\lambda_{\OH\OH \OO_{\Delta,J}}D_i \left({P_{24}\over P_{14}}\right)^{\Delta_{12}\over 2} {g^{(\Delta_{12},0)}_{\Delta,J}(u,v)\over P_{12}^{\Delta_1+\Delta_2\over 2}},
	\end{aligned}
\end{equation}
where the differential operators $D_i$ are given by \eqref{eq:DiffOp} and $\Delta_1=\Delta_2=d$.

The spinning correlator in embedding space with indices is then obtained using 
\begin{equation}\begin{aligned}
		\hat{G}_{MN,PS}(P_i)= {1\over 2^2({d\over 2}-1)^2}&\hat{D}^{(1)}_M \hat{D}^{(1)}_N\hat{D}^{(2)}_P \hat {D}^{(2)}_S \hat{G}(P_i,Z_i)
	\end{aligned}
\end{equation}
where $\hat{D}^{(i)}_M$ is given by 
\begin{equation}\label{eq:deffDhat}
	\hat{D}^{(i)}_M = \left({d-2\over 2}+Z_i\cdot {\partial\over \partial Z_i}\right){\partial \over \partial Z_i^M}-{1\over 2}Z_{i M} {\partial^2\over \partial Z^2}. 
\end{equation}
In order to project down to physical space we impose $P^M_{i} = (1,x_i^2,x_i^\mu)$ and contract  indices in embedding space with ${\partial P_i^M\over \partial x^\nu}=(0,2x^{(i)}_\nu,\delta^\mu_\nu)$ \cite{Costa:2011dw,Costa:2011mg}. We then set $x_1^\mu = (1,\vec{0})$, $x_2^{\mu}=(1+t,\vec{x})$, $x_3^\mu =(0,\vec{0})$ and $x_4\to \infty$ with $|x_{21}|\ll 1$ in the OPE limit, such that $u\to 0$ and $v\to 1$. 

\subsection{Stress tensor block}\label{sec:T}
The relations between different bases for the stress tensor three-point function can be found in e.g.\ Appendix C.1 in \cite{Hofman:2016awc}, some of which we summarize here for convenience. In the embedding space formalism \cite{Costa:2011mg,Costa:2011dw} the stress tensor three-point function can be built from \eqref{eq:GenthreePt}
\begin{equation}\label{eq:HVbasis}
	\langle T(P_1,Z_1)T(P_2,Z_2)T(P_3,Z_3)\rangle = {\sum_{p=1}^8 x_{p} Q_p\over  P_{12}^{d+2\over 2}P_{23}^{d+2\over 2}P_{31}^{d+2\over 2}},
\end{equation}
where the coefficients $x_p\equiv x_p^{(TTT)}$ are constrained by permutation symmetry and conservation to satisfy
\begin{equation}\label{eq:xp}\begin{aligned}
		x_1 &= 2x_2+{1\over 4}(d^2+2d-8)x_4-{1\over 2}d(d+2)x_7,\\
		x_8 &={1\over {d^2\over 2}-2}\Big[x_2-\Big({d\over 2}+1\Big)x_4+2dx_7\Big],\\
		x_2 &= x_3,\\
		x_4 &= x_5,\\
		x_6 &= x_7.
\end{aligned}\end{equation}

The stress tensor three-point function can be parameterized in terms of $(\hat{a},\hat{b},\hat{c})$ \cite{Osborn:1993cr} where one of these can further be traded for $C_T$ using the Ward identity 
\begin{equation}\label{eq:WardTTT}
	C_T = 4S_d {(d-2)(d+3)\hat{a}-2\hat{b}-(d+1)\hat{c}\over d(d+2)}.
\end{equation}
For the relation between the $x_p$ basis, $(\hat{a},\hat{b},\hat{c})$ and the $(t_2,t_4)$ coefficients that are natural when considering a conformal collider setup, we refer the reader to App.\ C in \cite{Hofman:2016awc}. However, we recall Eq.\ (C.10) in \cite{Hofman:2008ar} that relates these to $t_2$ and $t_4$ in $d=4$:
\begin{equation}
	\begin{aligned}
		t_2=& \frac{30(13\hat{a}+4\hat{b}-3\hat{c})}{14\hat{a}-2\hat{b}-5\hat{c}}\\
		t_4=& -\frac{15(81\hat{a}+32\hat{b}-20\hat{c})}{2(14\hat{a}-2\hat{b}-5\hat{c})}
	\end{aligned}
\end{equation}
and for $t_2=t_4=0$ one finds $\hat{a}={4\hat{c}\over 23}$ and $\hat{b}={17\hat{c}\over 92}$. On the other hand, the ratio of the anomaly coefficients $a,c$ is given by Eq.\ (C.12) in \cite{Hofman:2008ar}
\begin{equation}
	{a\over c} = {9\hat{a}-2\hat{b}-10\hat{c}\over 3(14\hat{a}-2\hat{b}-5\hat{c})}, 
\end{equation}
with $a=c$ when $t_2=t_4=0$. We further need the stress tensor three-point function with two heavy scalar operators
\begin{equation}
	\langle \OH(x_1)\OH(x_2)T_{\mu\nu}(x_3)\rangle = \lambda_{\OH\OH T_{\mu\nu}}\frac{W_\mu W_\nu-{1\over d}W^2\delta_{\mu\nu}}{ x_{12}^{2\Delta_H-2}x_{23}^{2}x_{31}^{2}},
\end{equation}
where $W^\mu = {x_{13}^\mu\over x_{13}^2}-{x_{23}^\mu\over x_{23}^2}$. Conformal Ward identities fix $\lambda_{\OH\OH T}$ to be
\begin{equation}
	\lambda_{\OH\OH T_{\mu\nu}} = -\frac{d}{d-1}{\Delta_H\over S_d},
\end{equation}
where $S_{d}={2\pi^{d\over 2}\over \Gamma({d\over 2})}$, which is related to $\mu$ and $\beta$ according to \eqref{eq:Temp} and \eqref{eq:defMu}.

From now on we consider $d=4$. For the stress tensor block we work with parametrization in terms of $(\hat{a},\hat{b},\hat{c})$.  In the channel $\hat{G}_{xy,xy}$, following the procedure described above, we obtain
\begin{equation}
	\begin{aligned}
		&\hat{G}_{xy,xy}|_T=\frac{\Delta_H }{2\pi^4(14 \hat{a}-2 \hat{b}-5 \hat{c})(t^2+\vec{x}^2)^5} \times\\
		&\times\Big[4 \hat{b} (-5 t^4 (x^2+y^2)+\vec{x}^2(x^4+6 x^2 y^2+y^4+(x^2+y^2) z^2)-4t^2 (x^4+9 x^2 y^2+y^4\\
		&+(x^2+y^2) z^2))+\hat{c} (-3 t^6+t^4 (13
		(x^2+y^2)-5 z^2)-\vec{x}^2(5 x^4-6 x^2 y^2+5 y^4\\
		&+4 (x^2+y^2) z^2-z^4)+t^2 (11 x^4+102
		x^2 y^2+11 y^4+10 (x^2+y^2) z^2-z^4))\\
		&+4 \hat{a} (t^6+t^4 (-17 (x^2+y^2)+3 z^2)-t^2 (13 x^4+106 x^2
		y^2+13 y^4+10 (x^2+y^2) z^2\\
		&-3 z^4)+\vec{x}^2(5 x^4-6 x^2 y^2+5 y^4+6 (x^2+y^2) z^2+z^4))\Big],
	\end{aligned}
\end{equation}
where $\vec{x}=(x,y,z)$ which after integrating over $x$ and $y$ gives 
\begin{equation}\label{eq:TxyxyInt}
	\begin{aligned}
		G_{xy,xy}|_T=\int dx dy \hat{G}_{xy,xy}|_T=-\frac{2 (7 \hat{a}+2 \hat{b}-\hat{c})\Delta_H (t^2-z^2)}{3\pi^3(14 \hat{a}-2 \hat{b}-5 \hat{c})(t^2+z^2)^2}.
	\end{aligned}
\end{equation}
and for $t_2=t_4=0$:
\begin{equation}\label{eq:TxyxyIntEinstein}
	G_{xy,xy}|_{T} = {2\Delta_H (t^2-z^2)\over 15\pi^3 (t^2+z^2)^2}.
\end{equation}
The results \eqref{eq:TxyxyInt} and \eqref{eq:TxyxyIntEinstein} are in agreement with \cite{Kulaxizi:2010jt}, where the explicit OPE was used to evaluate the stress tensor two-point function in a thermal state. 

Next, we find that $\hat{G}_{tx,tx}|_T$ is given by 
\begin{equation}\label{eq:Ttxtx}
	\begin{aligned}
		\hat{G}_{tx,tx}|_T=&\frac{\Delta_H }{2\pi^4(14 \hat{a}-2 \hat{b}-5 \hat{c})(t^2+\vec{x}^2)^5}\times\\
		&\times\Big[4\hat{b}(-t^6-x^2\vec{x}^4+t^4(-23x^2+4(y^2+z^2))\\
		&+t^2\vec{x}^2(9x^2+5(y^2+z^2)))+\hat{c}(15t^6-(5x^2-y^2-z^2)\vec{x}^4+t^4(41x^2\\
		&+7(y^2+z^2))-t^2\vec{x}^2(43x^2+7(y^2+z^2)))\\
		&+4\hat{a}(-13t^6+(3x^2-y^2-z^2)\vec{x}^4-3t^4(13x^2+y^2+z^2)+t^2\vec{x}^2(41x^2\\
		&+9(y^2+z^2)))\Big],
	\end{aligned}
\end{equation}
which becomes 
\begin{equation}
	\begin{aligned}
		G_{tx,tx}|_T=-\Delta_H{(64\hat{a}+14\hat{b}-19\hat{c})t^4-12(16\hat{a}+5\hat{b}-4\hat{c})t^2z^2+3(2\hat{b}+\hat{c})z^4\over 12(14\hat{a}-2\hat{b}-5\hat{c})\pi^3(t^2+z^2)^3},
	\end{aligned}
\end{equation}
after integrating over $x$ and $y$.
For $t_2=t_4=0$ this reduces to
\begin{equation}\label{eq:TtxtxIntEinstein}
	G_{tx,tx}|_T= \Delta_H{-9t^4+6t^2z^2+7z^4\over 60\pi^3 (t^2+z^2)^3}.
\end{equation}

Lastly, we consider $\hat{G}_{tz,tz}|_{T}$. Prior to integration, there is a $SO(3)$ rotational symmetry which allows us to obtain $\hat{G}_{tz,tz}|_{T}$ from \eqref{eq:Ttxtx} by $x\leftrightarrow z$. Integrating over the $xy$-plane we find
\begin{equation}
	\begin{aligned}
		G_{tz,tz}|_T={\Delta_H\over(14\hat{a}-2\hat{b}-5\hat{c})\pi^3(t^2+z^2)^4 }\Big[(-6\hat{a}+\hat{b}+2\hat{c})t^6+(-10\hat{a}-7\hat{b}+3\hat{c})t^4z^2\\
		+(30\hat{a}+7\hat{b}-8\hat{c})t^2z^4+(2\hat{a}-\hat{b}-\hat{c})z^6\Big].
	\end{aligned}
\end{equation}
For $t_2=t_4=0$ this reduces to
\begin{equation}\label{eq:TtztzIntEinstein}
	G_{tz,tz}|_T= \Delta_H{-105t^6+3t^4z^2+137t^2z^4+77z^6\over 270\pi^3 (t^2+z^2)^4}.
\end{equation}

\subsection{Spin-$0$ double-stress tensor block}
The simplest double-stress tensor operator is the scalar $[T^2]_{J=0}$ with dimension $\Delta_0$. The differential operators from the differential basis relevant here are $D_1$, $D_2$ and $D_6$ from \eqref{eq:DiffOp}, while the three-point function is given by \eqref{eq:SpinningThreePt} with $\lambda_{i,0}\equiv\lambda^{(i)}_{TT[T^2]_{J=0}}$. In order to impose conservation we demand that ${\partial\over \partial P_M}\hat{D}_M$ acting on \eqref{eq:SpinningThreePt} is $0$ \cite{Costa:2011mg}, where $\hat{D}_M$ is given by \eqref{eq:deffDhat}. This implies that the number of structures is reduced to just one
\begin{equation}\label{eq:consSpin0}
	\begin{aligned}
		\lambda_{2,0} &= -\frac{3}{4} (\Delta_0 -6) (\Delta_0 +2) \lambda_{1,0},\\
		\lambda_{6,0} &= \frac{3}{32} (\Delta_0 -6) (\Delta_0 -4) \Delta_0  (\Delta_0 +2)\lambda_{1,0},
	\end{aligned}
\end{equation}
and we are left with a single coefficient $\lambda_{1,0}$. The corresponding contribution to the correlator $\hat{G}(P_i,Z_i)$ (in embedding space) is given by 
\begin{equation}\label{eq:SpinningBlocksSpin0}
	\begin{aligned}
		\hat{G}(P_i,Z_i)|&_{[T^2]_0} =\sum_{i=1,3,6}\rho_{i,0}D_i W_{[T^2]_0},
	\end{aligned}
\end{equation}
where the conformal partial wave $W_{[T^2]_0}$ is given by \eqref{eq:PartialWave}. Note that the coefficients $\rho_{i,0}$ are related to $\lambda_{i,0}$ by an overall factor of the one-point function in the scalar state. They therefore, satisfy the same conservation condition as the $\lambda$'s in \eqref{eq:consSpin0}. The projection to the physical space and the relevant kinematics are described in the first part of this appendix.

\subsection{Spin-$2$ double-stress tensor block}
Because the spin-$2$ double-stress tensor $[T^2]_{J=2}$ is not conserved  there will be only two structures in the three-point function as compared to three for the stress tensor, even though they both have $J=2$. In the differential basis these can be labeled $\lambda_{i,2}\equiv \lambda^{(i)}_{TT[T^2]_{J=2}}$ with $i=1,2,\ldots 8$ in  \eqref{eq:SpinningThreePt}, which are reduced to two coefficients by imposing conservation:

\begin{align}
	\lambda_{3,2} &=\frac{\left(\Delta _2+2\right) \left(192 \lambda _{2,2}-\left(\Delta _2-4\right) \Delta _2 \left(\left(3 \Delta _2-16\right) \left(3 \Delta _2+4\right) \lambda _{1,2}+20 \lambda
		_{2,2}\right)\right)}{6 \Delta _2 \left(\Delta _2 \left(\Delta _2 \left(\left(\Delta _2-8\right) \Delta _2+2\right)+56\right)+96\right)},\nonumber\\
	\lambda_{4,2} &={\left(\Delta _2-4\right) \left(\Delta _2+2\right) \over 16 \left(\Delta _2 \left(\Delta _2 \left(\left(\Delta _2-8\right) \Delta
		_2+2\right)+56\right)+96\right)}\Big[(((\Delta _2-4) \Delta _2 (3 (\Delta _2-4) \Delta _2\nonumber\\
	&-52)-64) \lambda
	_{1,2}+4 (\Delta _2-8) (\Delta _2+4) \lambda _{2,2})\Big],\nonumber\\
	\lambda_{5,2} &=\frac{\left(\Delta _2-4\right) \Delta _2 \left(\left(15 \left(\Delta _2-4\right) \Delta _2+52\right) \lambda _{1,2}+52 \lambda _{2,2}\right)-96 \lambda _{2,2}}{12 \left(\Delta _2
		\left(\Delta _2 \left(\left(\Delta _2-8\right) \Delta _2+2\right)+56\right)+96\right)},\nonumber\\
	\lambda_{6,2} &={\left(\Delta _2-4\right) \Delta _2 \over 128 \left(\Delta _2 \left(\Delta _2 \left(\left(\Delta _2-8\right) \Delta _2+2\right)+56\right)+96\right)}\Big[((256\nonumber\\
	&-(\Delta _2-4) \Delta _2 ((\Delta _2-4) \Delta _2 (3 (\Delta _2-4)
	\Delta _2-56)+688)) \lambda _{1,2}\label{eq:consSpin2}\\
	&-4 ((\Delta _2-4) \Delta _2 (5 (\Delta _2-4) \Delta _2-52)+416) \lambda
	_{2,2})\Big]\nonumber\\
	&-\frac{48 \lambda _{2,2}}{\Delta _2 (\Delta _2 ((\Delta _2-8) \Delta
		_2+2)+56)+96},\nonumber\\
	\lambda_{7,2} &={\left(\Delta _2+2\right) \left(\Delta _2+4\right) \over 12 \left(\Delta _2-2\right) \Delta _2 \left(\Delta _2 \left(\Delta _2 \left(\left(\Delta _2-8\right) \Delta
		_2+2\right)+56\right)+96\right)}\Big[((\Delta _2-4) \Delta _2 \times\nonumber\\
	&\times((3 (\Delta _2-4) \Delta _2-44) \lambda _{1,2}+4
	\lambda _{2,2})-96 \lambda _{2,2})\Big],\nonumber\\
	\lambda_{8,2} &=-{\left(3 \left(\Delta _2-4\right) \Delta _2+16\right) \over 48 \left(\Delta _2 \left(\Delta _2 \left(\left(\Delta _2-8\right) \Delta _2+2\right)+56\right)+96\right)}\Big[((\Delta _2-4) \Delta _2 ((3 (\Delta _2-4) \Delta _2\nonumber\\
	&-44) \lambda
	_{1,2}+4 \lambda _{2,2})-96 \lambda _{2,2})\Big].\nonumber
\end{align}

The corresponding contribution to the correlator $\hat{G}(P_i,Z_i)$ is given by 
\begin{equation}\label{eq:SpinningBlocksSpin0secondformula}
	\begin{aligned}
		\hat{G}(P_i,Z_i)|&_{[T^2]_2} =\sum_{i=1}^8\rho_{i,2}D_i W_{[T^2]_2},
	\end{aligned}
\end{equation}
where the conformal partial wave $W_{[T^2]_2}$ is given by \eqref{eq:PartialWave}. Again, the coefficients $\rho_{i,2}$ are related to $\lambda_{i,2}$ by an overall factor of the one-point function in the scalar state. They therefore, satisfy the same conservation condition as the $\lambda$'s in \eqref{eq:consSpin2}. The projection to the physical space and the relevant kinematics are described in the first part of this appendix.

\subsection{Spin-$4$ double-stress tensor block}
For the spin-$4$ double-stress tensor operator $[T^2]_{J=4}$ there are a priori $10$ structures labelled by $\lambda_{i,4}\equiv \lambda^{(i)}_{TT[T^2]_{J=4}}$ with $i=1,2,\ldots 10$ in  \eqref{eq:SpinningThreePt}. Conservation reduces the number of structures to $3$ as follows:
{
	\allowdisplaybreaks
	\begin{align}
		\lambda_{4,4} &= {1\over96 ((\Delta _4-4) \Delta _4
			((\Delta _4-4) \Delta _4-44)+192)}\Big[(-((\Delta _4-4) \Delta _4 ((\Delta _4-4)\times\nonumber\\
		&\times \Delta _4 ((\Delta _4-4) \Delta _4 (3 (\Delta _4-4)
		\Delta _4-200)+5712)-92032))-485376) \lambda _{3,4}\nonumber\\
		&-2 (\Delta _4-6) (\Delta _4+4) (((\Delta _4-4)
		\Delta _4 ((\Delta _4-4) \Delta _4 (3 (\Delta _4-4) \Delta _4-68)-1024)\nonumber\\
		&+13056) \lambda _{1,4}+2 ((\Delta
		_4-4) \Delta _4 (3 (\Delta _4-4) \Delta _4-116)+768) \lambda _{2,4})\Big],\nonumber\\
		\lambda_{5,4} &= \frac{1}{8} (2 ((\Delta _4-4) \Delta _4+16) \lambda _{1,4}+(\Delta _4-8) (\Delta _4+2) \lambda _{3,4}+4 \lambda _{2,4}),\nonumber\\
		\lambda_{6,4} &={1\over256 (\Delta _4-6) (\Delta _4+4)
			((\Delta _4-4) \Delta _4 ((\Delta _4-4) \Delta _4-44)+192)}\Big[2 (\Delta _4-6)\times\nonumber\\
		&\times (\Delta _4+4) (((\Delta _4-4) \Delta _4 ((\Delta _4-4) \Delta _4 ((\Delta
		_4-4) \Delta _4 ((\Delta _4-4) \Delta _4 ((\Delta _4-4) \Delta _4\nonumber\\
		&-64)+1040)+11392)-262144)+2162688)
		\lambda _{1,4}+2 ((\Delta _4-4) \Delta _4\times\nonumber\\
		&\times((\Delta _4-4) \Delta _4 ((\Delta _4-4) \Delta _4 ((\Delta _4-4)
		\Delta _4-88)+2448)-17408)\nonumber\\
		&+86016) \lambda _{2,4})+((\Delta _4-4) \Delta _4 ((\Delta _4-4) \Delta _4
		((\Delta _4-4) \Delta _4\times\nonumber\\
		&\times ((\Delta _4-4) \Delta _4 ((\Delta _4-4) \Delta _4 ((\Delta _4-4) \Delta
		_4-108)+5104)-131904)\nonumber\\
		&+2009088)-18300928)+81788928) \lambda _{3,4}\Big],\nonumber\\
		\lambda_{7,4} &= {1\over 24
			((\Delta _4-4) \Delta _4 ((\Delta _4-4) \Delta _4-44)+192)}\Big[(\Delta _4-4) (\Delta _4+6) \times\nonumber\\
		&\times(2 (\Delta _4-6) (\Delta _4+4) (((\Delta _4-4) \Delta _4+20)
		\lambda _{1,4}+2 \lambda _{2,4})+((\Delta _4-4) \Delta _4\times\label{eq:consSpin4}\\
		&\times ((\Delta _4-4) \Delta _4-36)+704) \lambda _{3,4})\Big],\nonumber\\
		\lambda_{8,4} &= {1\over 32 (\Delta _4-6) (\Delta _4+4)
			((\Delta _4-4) \Delta _4 ((\Delta _4-4) \Delta _4-44)+192)}\Big[2 (\Delta _4-6)\times\nonumber\\
		&\times (\Delta _4+4) (((\Delta _4-4) \Delta _4 ((\Delta _4-4) \Delta _4 ((\Delta
		_4-4) \Delta _4-20) ((\Delta _4-4) \Delta _4\nonumber\\
		&+24)+4736)+135168) \lambda _{1,4}+2 ((\Delta _4-4) \Delta _4
		((\Delta _4-6) (\Delta _4-4) \Delta _4 (\Delta _4+2)\nonumber\\
		&-320)+7680) \lambda _{2,4})+(\Delta _4 (\Delta _4
		(\Delta _4 (\Delta _4 (\Delta _4 (((\Delta _4-20) \Delta _4+120) \Delta _4^3-1968 \Delta
		_4\nonumber\\
		&+2112)+23296)-78848)-327680)+1638400)+5111808) \lambda _{3,4}\Big],\nonumber\\
		\lambda_{9,4} &={1\over 3 (\Delta _4-6) (\Delta _4+4) ((\Delta
			_4-4) \Delta _4 ((\Delta _4-4) \Delta _4-44)+192)}\Big[(\Delta _4-6)\times\nonumber\\
		&\times (\Delta _4+4)\times (((\Delta _4-4) \Delta _4 (9 (\Delta _4-4) \Delta _4+68)-768)
		\lambda _{1,4}\nonumber\\
		&+16 ((\Delta _4-4) \Delta _4-6) \lambda _{2,4})+((\Delta _4-4) \Delta _4 ((\Delta _4-4) \Delta _4\times\nonumber\\
		&\times(3 (\Delta _4-4) \Delta _4-116)+3136)-15360) \lambda _{3,4}\Big],\nonumber\\
		\lambda_{10,4} &={1\over 12 ((\Delta _4-4) \Delta _4 ((\Delta _4-4) \Delta _4-44)+192)}\Big[((\Delta _4-4) \Delta _4+12)\times\nonumber\\
		&\times(-2 (\Delta _4-6)\times (\Delta _4+4) (((\Delta _4-4) \Delta
		_4+20) \lambda _{1,4}+2 \lambda _{2,4})\nonumber\\
		&-((\Delta _4-4) \Delta _4 ((\Delta _4-4) \Delta _4-36)+704) \lambda
		_{3,4})\Big].\nonumber
	\end{align}
}

The corresponding contribution to the correlator $\hat{G}(P_i,Z_i)$ is given by 
\begin{equation}\label{eq:SpinningBlocksSpin4}
	\begin{aligned}
		\hat{G}(P_i,Z_i)|&_{[T^2]_4} =\sum_{i=1}^{10}\rho_{i,4}D_i W_{[T^2]_4},
	\end{aligned}
\end{equation}
where the conformal partial wave $W_{[T^2]_4}$ is given by \eqref{eq:PartialWave}. The coefficients $\rho_{i,4}$ are related to $\lambda_{i,4}$ by an overall factor of the one-point function in the scalar state. They therefore, satisfy the same conservation condition as the $\lambda$'s in \eqref{eq:consSpin4}. The projection to the physical space and the relevant kinematics are described in the first part of this appendix.

\subsection{Integrated double stress tensor contribution}
In this section we list the explicit expression for the integrated $\OO(C_T\mu^2)$ part of the conformal block expansion of $G_{xy,xy}$, $G_{tx,tx}$ and $G_{tz,tz}$ obtained using the procedure described above. We regulate divergences by including a factor of $|x|^{-\epsilon}$, which produces simple poles as $\epsilon\to0$.  For $G_{xy,xy}$ we find as $\epsilon\to0$:
\begin{equation}\label{eq:xyxyCFT}
	G_{xy,xy}|_{\mu^2 C_T}  = p^{(0)}_{xy,xy}(t,z)+p^{(1)}_{xy,xy}(t,z)\log(t^2+z^2)+{c_1 t^2+c_2 z^2\over \epsilon}
\end{equation}
where $c_1,c_2$ are some constants depending on the CFT data and
\begin{equation}
	p^{(0)}_{xy,xy}(t,z) = \frac{\pi ^5 \mu ^2 C_T}{1693440000(t^2+z^2)}\sum_{j=0}^2 p^{(0,2j)}_{xy,xy}t^{4-2j}z^{2j}
\end{equation}
with 
\begin{equation}
	\begin{aligned}
		p^{(0,0)}_{xy,xy}&=-8 (22050 \rho _{1,0}^{(1)}-162243 \rho _{1,2}^{(1)}-11683490 \rho _{1,4}^{(1)}+129168 \rho _{2,2}^{(1)}+4702775 \rho _{2,4}^{(1)}\\
		&+6991740 \rho _{3,4}^{(1)})+4410 \gamma^{(1)}
		_0+304479 \gamma^{(1)} _2-3577875 \gamma^{(1)} _4,\\
		p^{(0,2)}_{xy,xy}&=2\Big(-8(22050 \rho _{1,0}^{(1)}-89343 \rho _{1,2}^{(1)}-3641540 \rho _{1,4}^{(1)}+56268 \rho _{2,2}^{(1)}+1646645 \rho _{2,4}^{(1)}\\
		&+2005920 \rho _{3,4}^{(1)})+4410 \gamma^{(1)}
		_0+14364 \gamma^{(1)} _2-964005 \gamma^{(1)} _4\Big),\\
		p^{(0,4)}_{xy,xy}&=7\Big(8 (-3150 \rho _{1,0}^{(1)}+2349 \rho _{1,2}^{(1)}-215350 \rho _{1,4}^{(1)}+2376 \rho _{2,2}^{(1)}+90475 \rho _{2,4}^{(1)}\\
		&+123300 \rho _{3,4}^{(1)})+630 \gamma^{(1)} _0-39393 \gamma^{(1)}
		_2+74415 \gamma^{(1)} _4\Big),\\
	\end{aligned}
\end{equation}
and 
\begin{equation}
	p^{(1)}_{xy,xy}(t,z) = -\frac{\pi ^5 \mu ^2 C_T}{15680000}\sum_{j=0}^1 p^{(1,2j)}_{xy,xy}t^{2-2j}z^{2j}
\end{equation}
with
\begin{equation}
	\begin{aligned}
		p^{(1,0)}_{xy,xy}&=3\Big(16 (-702 \rho _{1,2}^{(1)}-19565 \rho _{1,4}^{(1)}+702 \rho _{2,2}^{(1)}+8085 \rho _{2,4}^{(1)}\\
		&+11480 \rho _{3,4}^{(1)})+490 \gamma^{(1)} _0-5607 \gamma^{(1)} _2+12040 \gamma^{(1)}
		_4\Big),\\
		p^{(1,2)}_{xy,xy}&=\Big(16 (486 \rho _{1,2}^{(1)}-6055 \rho _{1,4}^{(1)}-486 \rho _{2,2}^{(1)}\\
		&+2695 \rho _{2,4}^{(1)}+3360 \rho _{3,4}^{(1)}+280 \gamma^{(1)} _4)+1470 \gamma^{(1)} _0-189 \gamma^{(1)} _2\Big).
	\end{aligned}
\end{equation}

Next, for $(\partial_t^2+\partial_z^2)G_{tx,tx}$ we find
\begin{equation}\label{eq:txtxCFT}
	(\partial_t^2+\partial_z^2)G_{tx,tx}|_{\mu^2 C_T}  = p^{(0)}_{tx,tx}(t,z)+p^{(1)}_{tx,tx}\log(t^2+z^2)+{c_3\over \epsilon}
\end{equation}
for some constant $c_3$ where
\begin{equation}
	p^{(0)}_{tx,tx}(t,z) = -\frac{\pi ^5 \mu ^2 C_T}{423360000 \left(t^2+z^2\right)^3}\sum_{j=0}^3 p^{(0,2j)}_{tx,tx}t^{6-2j}z^{2j}
\end{equation}
with 
\begin{equation}
	\begin{aligned}
		p^{(0,0)}_{tx,tx}&=176400 \rho _{1,0}^{(1)}-265032 \rho _{1,2}^{(1)}+30139760 \rho _{1,4}^{(1)}+529632 \rho _{2,2}^{(1)}-12698840 \rho _{2,4}^{(1)}\\
		&-17881920 \rho _{3,4}^{(1)}+97020 \gamma^{(1)} _0-345492 \gamma^{(1)}
		_2-792435 \gamma^{(1)} _4,\\
		p^{(0,2)}_{tx,tx}&=\frac{1}{48} (25401600 \rho _{1,0}^{(1)}+203700096 \rho _{1,2}^{(1)}-9623496960 \rho _{1,4}^{(1)}-165597696 \rho _{2,2}^{(1)}\\
		&+3983253120 \rho _{2,4}^{(1)}+5576739840 \rho
		_{3,4}^{(1)}+21591360 \gamma^{(1)} _0+81539136 \gamma^{(1)} _2\\
		&+356907600 \gamma^{(1)} _4),\\
		p^{(0,4)}_{tx,tx}&=\frac{1}{48} (25401600 \rho _{1,0}^{(1)}+270884736 \rho _{1,2}^{(1)}+270063360 \rho _{1,4}^{(1)}-232782336 \rho _{2,2}^{(1)}\\
		&-40360320 \rho _{2,4}^{(1)}-293207040 \rho _{3,4}^{(1)}+29211840
		\gamma^{(1)} _0+125629056 \gamma^{(1)} _2\\
		&-45889200 \gamma^{(1)} _4),\\
		p^{(0,6)}_{tx,tx}&=176400 \rho _{1,0}^{(1)}+1134648 \rho _{1,2}^{(1)}-6309520 \rho _{1,4}^{(1)}-870048 \rho _{2,2}^{(1)}+2824360 \rho _{2,4}^{(1)}\\
		&+3044160 \rho _{3,4}^{(1)}+255780 \gamma^{(1)} _0+573048 \gamma^{(1)} _2-71715
		\gamma^{(1)} _4,
	\end{aligned}
\end{equation}
and 
\begin{equation}\begin{aligned}
		p^{(1)}_{tx,tx}&=-\frac{3 \pi ^5 \mu ^2 C_T }{15680000}\Big[-8 \left(4 \left(81 \rho _{1,2}^{(1)}+35 \rho _{1,4}^{(1)}-81 \rho _{2,2}^{(1)}-35 \rho _{3,4}^{(1)}\right)+315 \gamma^{(1)} _4\right)\\
		&+980 \gamma^{(1)} _0+63 \gamma^{(1)}
		_2\Big].
\end{aligned}\end{equation}

Lastly, for  $(\partial_t^2+\partial_z^2)^2G_{tz,tz}$ we find
\begin{equation}\label{eq:tztzCFT}
	(\partial_t^2+\partial_z^2)^2G_{tz,tz}|_{\mu^2 C_T}  = \frac{\pi ^5 \mu ^2 C_T}{2940000 \left(t^2+z^2\right)^5}\sum_{j=0}^4 p^{(0,2j)}_{tz,tz}t^{8-2j}z^{2j},
\end{equation}
where
\begin{equation}
	\begin{aligned}
		p^{(0,0)}_{tz,tz}&=-7776 \rho _{1,2}^{(1)}+197120 \rho _{1,4}^{(1)}+7776 \rho _{2,2}^{(1)}-86240 \rho _{2,4}^{(1)}-110880 \rho _{3,4}^{(1)}\\
		&+1470 \gamma^{(1)} _0+189 \gamma^{(1)} _2-1400 \gamma^{(1)} _4,\\
		p^{(0,2)}_{tz,tz}&=-248832 \rho _{1,2}^{(1)}+19983040 \rho _{1,4}^{(1)}+248832 \rho _{2,2}^{(1)}-8451520 \rho _{2,4}^{(1)}\\
		&-11531520 \rho _{3,4}^{(1)}-5880 \gamma^{(1)} _0-152712 \gamma^{(1)}_2-845320 \gamma^{(1)} _4,\\
		p^{(0,4)}_{tz,tz}&=233280 \rho _{1,2}^{(1)}-82577600 \rho _{1,4}^{(1)}-233280 \rho _{2,2}^{(1)}+34496000 \rho _{2,4}^{(1)}\\
		&+48081600 \rho _{3,4}^{(1)}-14700 \gamma^{(1)} _0+82530 \gamma^{(1)} _2+3193400 \gamma^{(1)} _4,\\
		p^{(0,6)}_{tz,tz}&=435456 \rho _{1,2}^{(1)}+29986880 \rho _{1,4}^{(1)}-435456 \rho _{2,2}^{(1)}-12246080 \rho _{2,4}^{(1)}\\
		&-17740800 \rho _{3,4}^{(1)}-5880 \gamma^{(1)} _0+218736 \gamma^{(1)} _2-1147160 \gamma^{(1)}_4,\\
		p^{(0,8)}_{tz,tz}&=-38880 \rho _{1,2}^{(1)}-257600 \rho _{1,4}^{(1)}+38880 \rho _{2,2}^{(1)}+86240 \rho _{2,4}^{(1)}+171360 \rho _{3,4}^{(1)}\\
		&+1470 \gamma^{(1)} _0-16695 \gamma^{(1)} _2+12320 \gamma^{(1)} _4.
	\end{aligned}
\end{equation}

\subsection{Comparison with the bulk calculations}
Solving \eqref{eq:matching} we find the anomalous dimensions \eqref{eq:solAnomDim}, the relations \eqref{eq:solCoeff} and the following bulk coefficients $(a^{(xy)}_{8,1}(6,0),a^{(xy)}_{8,1}(8,0),a^{(tx)}_{8,1}(8,0))$:
\begin{equation}\label{eq:solACoeff}
	\begin{aligned}
		a^{8,1(xy)}_{6,0} &= \frac{\pi ^4 \mu ^2 \left(2 \rho _{1,0}^{(1)}-3 \rho _{1,2}^{(1)}+\rho _{1,4}^{(1)}\right)}{1440}-\frac{3150449 \mu ^2}{47628000}+\frac{1441 \mu ^2}{37800 \epsilon },\\
		a^{8,1(xy)}_{8,0} &=-\frac{\pi ^4 \mu ^2 \left(2 \rho _{1,0}^{(1)}-3 \rho _{1,2}^{(1)}+\rho _{1,4}^{(1)}\right)}{1920}+\frac{1820863 \mu ^2}{127008000}-\frac{1801 \mu ^2}{50400 \epsilon },\\
		a^{8,1(tx)}_{8,0} &=\frac{\pi ^4 \mu ^2 \left(2 \rho _{1,0}^{(1)}+3 \rho _{1,2}^{(1)}-5 \rho _{1,4}^{(1)}\right)}{2880} -\frac{132403 \mu ^2}{1411200}-\frac{47 \mu ^2}{45360 \epsilon},
	\end{aligned}
\end{equation}
which are divergent as $\epsilon\to0$. Note that by also studying the $G_{xx,xx}$ polarization we get one more linearly independent equation:
\begin{equation}\label{eq:Axxxx}
	a^{8,1(xx)}_{0,0}=\frac{\pi ^3 \mu ^2 \left(\rho _{1,0}^{(1)}+2 \rho _{1,4}^{(1)}\right)}{80}-\frac{6713281 \mu ^2}{5292000 \pi }+\frac{11741 \mu ^2}{6300 \pi  \epsilon }.
\end{equation}

\bibliographystyle{JHEP}
\bibliography{refs} 

\end{document}